\documentclass[superscriptaddress,aps,twocolumn,preprintnumbers,showpacs,showkeys,nofootinbib,floatfix]{revtex4-1}

\usepackage{amssymb,amsmath,amsfonts,amsthm,graphicx,psfrag}

\newcommand{\beq}{\begin{eqnarray}}
\newcommand{\eeq}{\end{eqnarray}}

\newcommand{\tr}{\operatorname{Tr}}

\begin{document}
\preprint{}

\title{Lattice field theory simulations of Dirac semimetals}

\author{V.~V.~Braguta}
\email[]{braguta@itep.ru}
\affiliation{Institute for Theoretical and Experimental Physics, 117259 Moscow, Russia}
\affiliation{Institute for High Energy Physics NRC "Kurchatov Institute", Protvino, 142281 Russian Federation}
\affiliation{Far Eastern Federal University, School of Biomedicine, 690950 Vladivostok, Russia}
\affiliation{Moscow Institute of Physics and Technology, Institutskii per. 9, Dolgoprudny, Moscow Region, 141700 Russia}

\author{M.~I.~Katsnelson}
\email[]{m.katsnelson@science.ru.nl}
\affiliation{Radboud University, Institute for Molecules and Materials,
Heyendaalseweg 135, NL-6525AJ Nijmegen, The Netherlands}
\affiliation{Ural Federal University, Theoretical Physics and Applied Mathematics Department, Mira Str. 19, 620002 Ekaterinburg, Russia}

\author{A.~Yu.~Kotov}
\email[]{kotov@itep.ru}
\affiliation{Institute for Theoretical and Experimental Physics, 117259 Moscow, Russia}

\begin{abstract}
In this paper the observed Dirac semimetals Na$_3$Bi and Cd$_3$As$_2$ are studied within lattice simulation. 
We formulate lattice field theory with rooted staggered fermions on anisotropic lattice.
It is shown that in the limit of zero temporal lattice spacing this theory reproduces low energy effective theory 
of Dirac semimetals. Using this lattice theory
we study the phase diagram of Dirac semimetals in the plane effective coupling constant--Fermi velocity 
anisotropy. Within the formulated theory the results are practically volume independent
in contrast with our previous study. Our results confirm our previous finding
that within the Dirac model with bare Coulomb interaction both Na$_3$Bi and Cd$_3$As$_2$ 
lie deep in the insulator phase.
\end{abstract}

\keywords{Semimetal, insulator, Coulomb interaction, Monte-Carlo simulations}

\pacs{71.30.+h, 05.10.Ln}

\maketitle

\section{Introduction}

Recent years are marked with the discovery of numerous materials with interesting properties. 
Considerable interest is attracted by the materials which low-energy fermionic excitations can be described by massless fermions. Among well-known examples is graphene\cite{Novoselov666, Geim2007} - two-dimensional material with  two effective massless Dirac fermions\cite{PhysRev.71.622,PhysRev.104.666,Semenoff:1984dq,Novoselov2005,Zhang2005}. Its three-dimensional analogues include Dirac(Na$_3$Bi\cite{Liu864}, Cd$_3$As$_2$\cite{Neupane2014,PhysRevLett.113.027603}) and Weyl semimetals\cite{Xu613,Xue1501092}. Such materials provide a perfect opportunity for detailed study of quantum field theory phenomena, 
which were previously related to high-energy physics.

In this paper we are going to concentrate on studying Dirac semimetals. Low energy spectrum of 
the observed Dirac semimetals is determined by two Fermi points. In the vicinity of each Fermi point
the fermion excitations reveal the properties of massless 3D Dirac fermions with the dispersal relation
\begin{equation}
E^2=v^2_{\parallel}(k^2_x+k_y^2)+v^2_{\perp}k^2_z,
\label{eq:dispersion}
\end{equation}
where $v_{\parallel}, v_{\perp}$ are Fermi velocities in the $(x,y)$ plane and $z$ direction correspondingly.
For the Na$_3$Bi: $v_{\parallel}/c\simeq0.001, v_{\perp}/v_{\parallel}\simeq0.1$\cite{Liu864} and for the Cd$_3$As$_2$:
$v_{\parallel}/c \simeq0.004, v_{\perp}/v_{\parallel}\simeq0.25$\cite{Liu2014}, where $c$ is the light velocity (further we will express all velocities in the units of $c$). 

Electromagnetic interaction between quasiparticles might significantly modify the properties of these systems and thus give rise to a plenty of interesting phenomena\cite{RevModPhys.84.1067}. The reason for that is the smallness of Fermi velocity of quasiparticles $v_F\ll 1$. As the result the interaction is reduced to instantaneous Coulomb with effective coupling constant $\alpha_{eff}=\alpha \cdot 1 / v_F \sim 1$, where $\alpha\approx1/137$ is the coupling constant of electromagnetic interaction. For instance, the effective coupling constants 
can be estimated as $\alpha_{eff} \sim 7$ for the 
Na$_3$Bi and $\alpha_{eff} \sim 2$ for the Cd$_3$As$_2$. So, the coupling constants 
of both materials are large and they are strongly correlated.  It is known that large 
coupling constant can cause dynamical generation of the fermion mass gap
thus leading to the transition from semimetal to the insulator phase\cite{Khveshchenko:2001zz,Gamayun:2009em,Sabio:2010yf}. The study of this transition is 
interesting and important from theoretical and experimental points of view. 

Since Dirac semimetals are strongly correlated materials investigation of their properties requires nonperturbative consideration.
We believe that today the only way to carry out model independent study of strongly coupled systems is lattice Quantum Monte Carlo (QMC) simulation \cite{Montvay:1994cy}. 
This approach fully accounts many body effects in strongly coupled systems. Lattice simulation was intensively used to study the 
properties of graphene \cite{Drut:2008rg, Hands:2008id,Armour:2009vj, Drut:2009aj, Ulybyshev:2013swa, Boyda:2016emg}. Lattice 
simulation was also used to study topological phase transitions in Weyl semimetals\cite{Yamamoto:2016rfr, Yamamoto:2016zpx}. In this paper we are going to use 
lattice simulation to study semimetals/insulator transition in Dirac semimetals. 

The first lattice study of the semimetals/insulator transition in Dirac semimetals was carried out in paper \cite{Braguta:2016vhm}. 
In particular, in this paper dynamical chiral symmetry breaking caused by strong interaction was addressed. 
The result of this study allowed to find critical coupling constant for the semimetals/insulator transition
and draw the phase diagram in the plane $(\alpha_{eff}, v_{\perp}/v_{\parallel})$. 
The calculation done in \cite{Braguta:2016vhm} was carried out at two lattice volumes 
and exhaustive study of finite volume effects was not done. 

In this paper we continue the study of dynamical chiral symmetry breaking caused by strong interaction in Dirac semimetals 
by means of lattice simulation. We develop an approach that allows to carry out lattice simulation of Dirac semimetals 
keeping finite volume uncertainties under control. Using this approach 
we calculate critical coupling for the  semimetals/insulator transition as a function of the Fermi velocity 
asymmetry $v_{\perp}/v_{\parallel}$ and draw the phase diagram of Dirac semimetals. 

This paper is organized as follows. In the next section we formulate low energy 
effective field theory of Dirac semimetals in continuum. Section III
is devoted to the description of lattice field theory of Dirac semimetals.
Using this lattice theory we study the phase diagram of Dirac semimetals 
in section IV. In last section we discuss our results. 

\section{Effective field theory}
Low-energy fermionic excitations of the discovered semimetals can be described as $N_f=2$ massless 3D Dirac fermions. 
Their dispersion relation is given by formula (\ref{eq:dispersion}). Important features of the 
fermionic spectrum are the smallness of the Fermi velocity $v_F \ll 1$ and the anisotropy of the Fermi velocity in 
different directions $v_{\perp}/v_{\parallel}<1$. The smallness of the Fermi velocity allows us 
to state that magnetic interactions and retardation effects are considerably suppressed in Dirac semimetals. 
This implies that the electromagnetic interaction between quasiparticles is reduced to the instantaneous Coulomb potential. 

\begin{figure*}[t]
\begin{tabular}{cc}
\includegraphics[scale=0.45,clip=false]{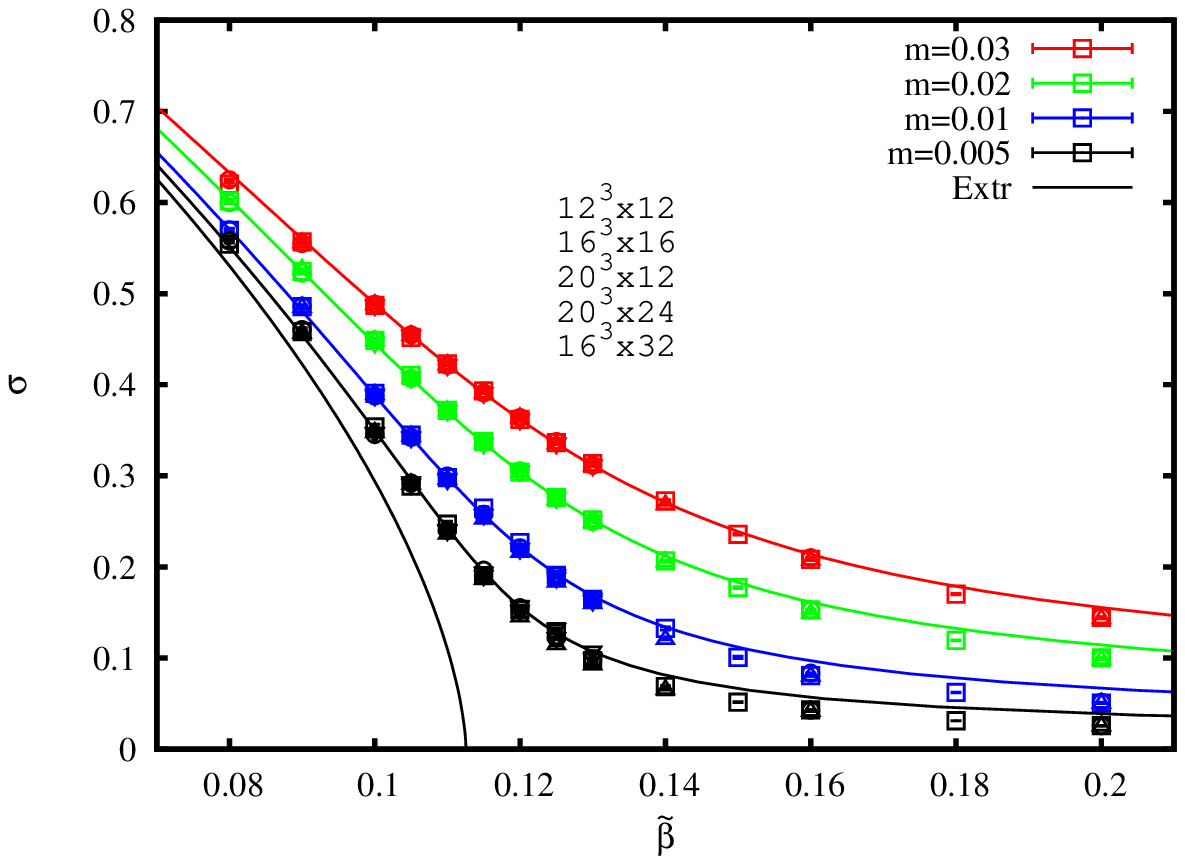} & 
\includegraphics[scale=0.45,clip=false]{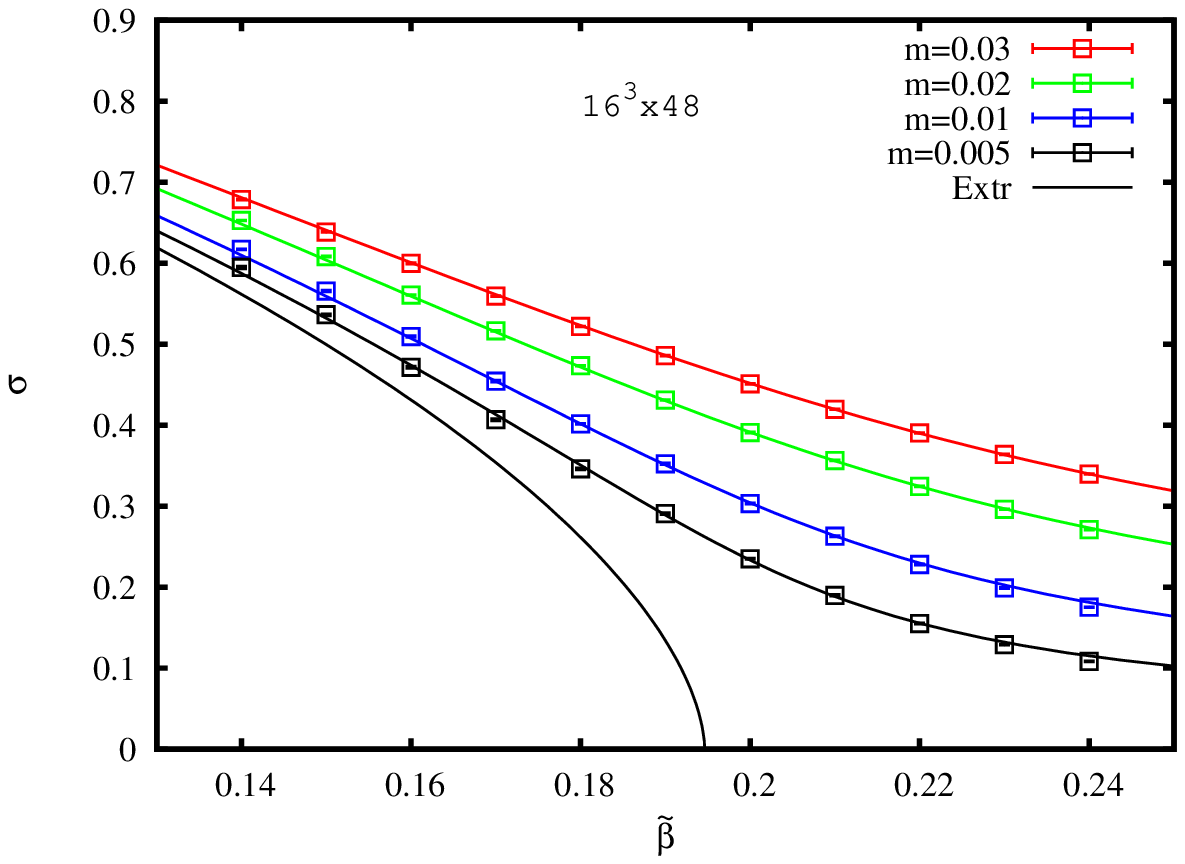} \\
$\xi=1/2$ & $\xi=1/3$ \\
\\
\includegraphics[scale=0.45,clip=false]{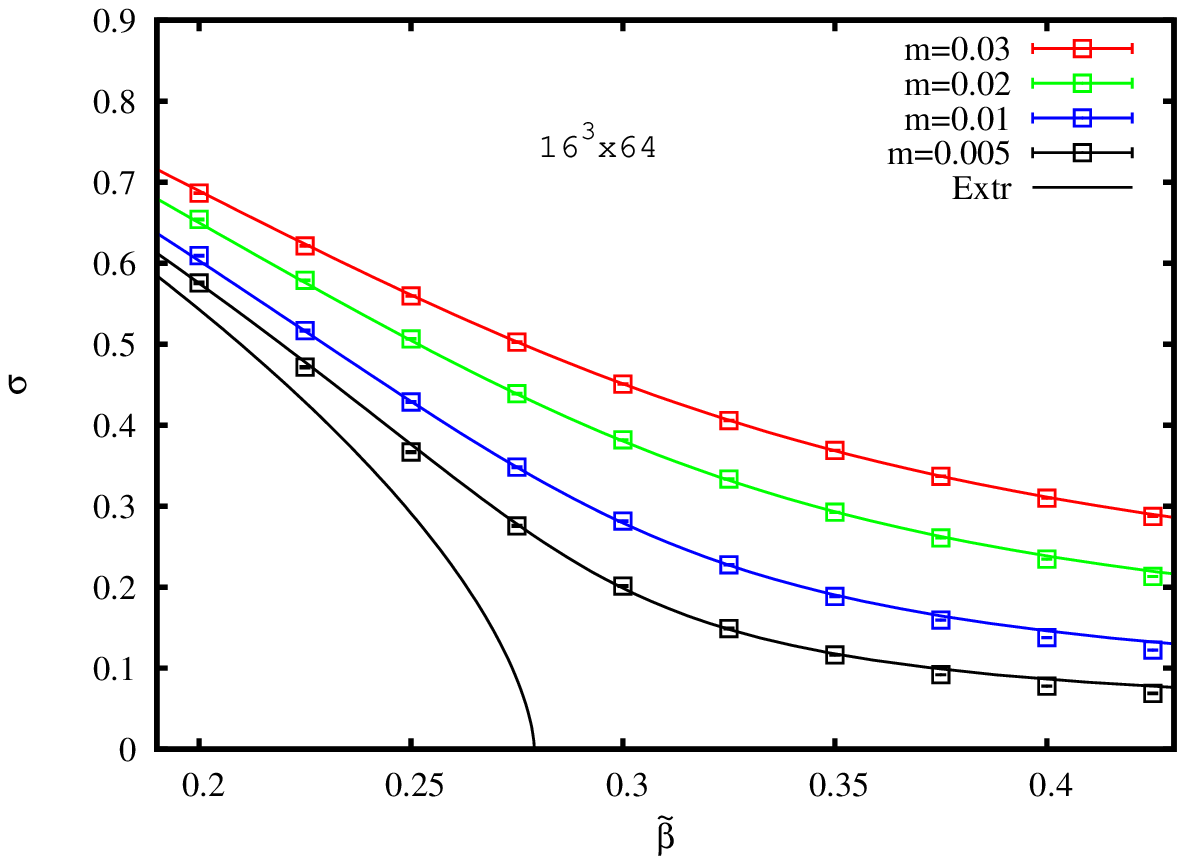} &
\includegraphics[scale=0.45,clip=false]{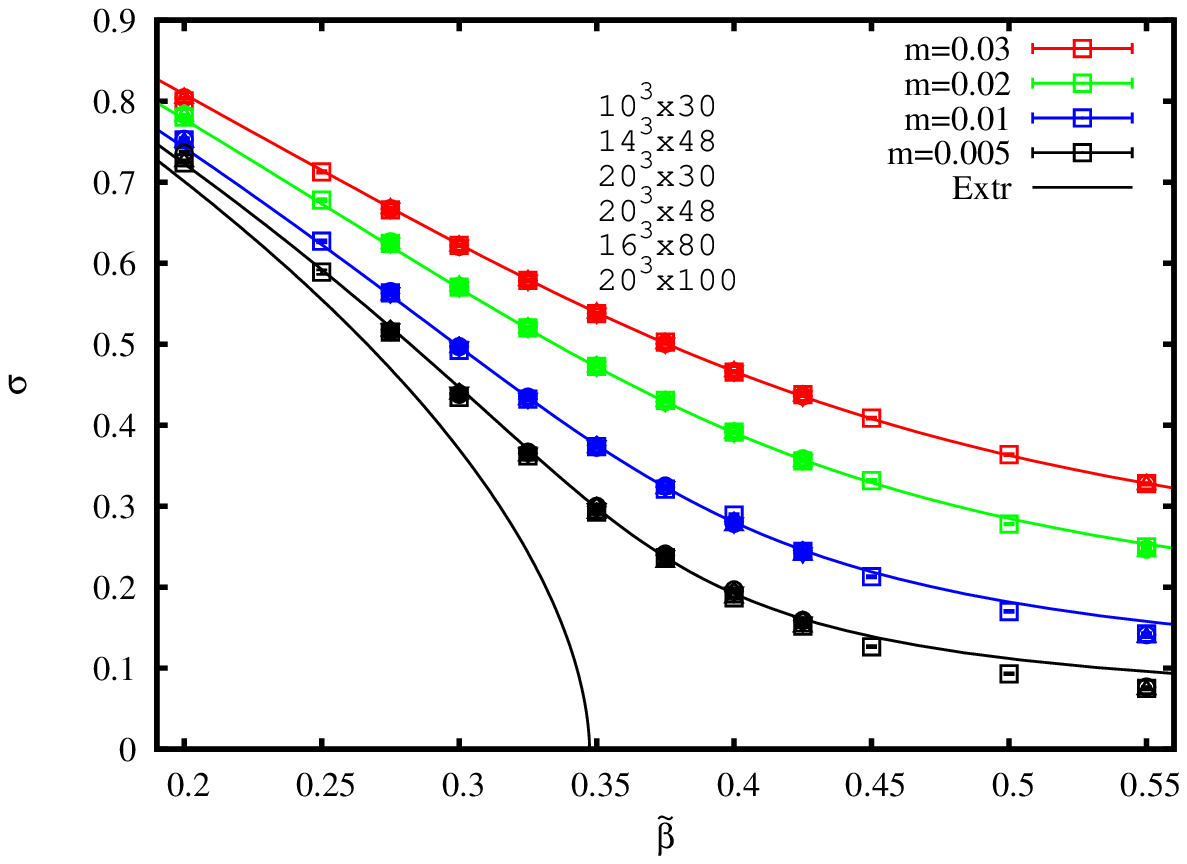} \\
 $\xi=1/4$ & $\xi=1/5$
\end{tabular}
\caption{The dependence of the chiral condensate on the $\tilde{\beta}$ (Eq. (\ref{eq:betadef})) for different values of $m$. The points for different lattice sizes almost coincide. Errorbars are smaller than data points. Black line corresponds to extrapolation to $m=0$ obtained with the help of EoS (\ref{eq:eos}). Different plots correspond to different values of $\xi=\frac{a_t}{a_s}$ (Eq. (\ref{eq:xidefinition})).
}
\label{fig:isopsipsi}
\end{figure*}

The properties described above allow one to build low energy effective quantum field theory of fermionic excitations in Dirac semimetals. In particular, the partition function of theory can be written as a path integral
\beq
Z=\int D\psi~ D\bar\psi~DA_4~\exp{\bigl ( -S_E \bigr )},
\label{Z}
\eeq
where $\bar \psi, \psi$ are fermion fields, $A_4$ is temporal component of the vector potential of the electromagnetic field.
The Euclidean action $S_E$ can be written as a sum of the contributions of the gauge fields $S_{g}$ and the fermions $S_f$ 
\beq
S_E=S_f+S_g.
\label{eq:continuousaction}
\eeq
The gauge field contribution can be written as 
\beq
S_g = \frac{1}{8\pi\alpha}\int d^3x dt (\partial_i A_4)^2,
\eeq
whereas the fermion contribution is 
\beq
S_f &=& \sum\limits_{a=1}^{N_f}\int d^3x dt \bar{\psi}_a \left(\gamma_4(\partial_4+iA_4)+v_{F,i}\gamma_i\partial_i\right)\psi_a  
\nonumber \\  
&=& \sum\limits_{a=1}^{N_f}\int d^3x dt \bar{\psi}_a D_a(A_4) \psi_a.
\label{sf}
\eeq
In last two equations the following designations are introduced: 
the $\alpha\approx1/137$ is the electromagnetic coupling constant, 
$v_{F,i}$ are the components of Fermi velocity: $v_{F,x}=v_{F,y}=v_{\parallel}$, $v_{F,z}=v_{\perp}$,  
the $\gamma_{\mu},~ \mu=1,2,3,4$ are the standard Dirac matrices. 
The $\psi_a$ is four component Dirac spinor,  the index $a$ counts the contributions of different Fermi points 
to the low energy effective action. For further use we also introduced the designation of the Dirac operator $D_a(A_4)$. 
The fermions in the effective theory (\ref{sf}) are degenerate what allows one 
to omit the Fermi point index $a$ of the Dirac operator.

Note that the action (\ref{eq:continuousaction}) includes only time-like component $A_4$ of the electromagnetic field. 
As was explained above this is because the contribution of space-like components is suppressed by the factors of 
$v_F$ and can be neglected due to the smallness of Fermi velocity. The action $S_E$ is quadratic in the $A_4$
so this field can be integrated out what leads to the Coulomb interaction between fermions 
$\frac{\rho(\vec{r})\rho(\vec{r'})}{|\vec{r}-\vec{r'}|}$, where $\rho(\vec{r})=\bar\psi_a(\vec{r})\gamma_4\psi_a(\vec{r})$ 
is the charge density. Thus the partition function (\ref{Z}) describes the system of $N_f$ Dirac fermions which 
interact through instantaneous Coulomb law.

Let us now rescale the time coordinate and timelike component of the electromagnetic field:
\beq
	t   \to & t\cdot v_F, \nonumber \\
	A_4 \to & A_4/v_F, 
\eeq
what brings us to the action 
\begin{equation}
\begin{split}
    S_E=\sum\limits_{a=1}^{N_f=2}\int d^3x dt \bar{\psi}_a \left(\gamma_4(\partial_4+iA_4)+\zeta_i \gamma_i\partial_i\right)\psi_a+\\
    +\frac{1}{8\pi\alpha_{eff}}\int d^3x dt (\partial_i A_4)^2.
\label{eq:continuousactionscaled}
\end{split}
\end{equation}
In the above equation we introduced Fermi velocity anisotropy factors $\zeta_i=\frac{v_{F,i}}{v_F}$.
It is seen from equation (\ref{eq:continuousactionscaled}) that the strength of the interaction is determined by the effective coupling constant $\alpha_{eff}=\frac{\alpha}{v_F}$. 
For the observed Dirac semimetals the effective coupling constants can be estimated as $\alpha_{eff} \approx 7$ for the 
Na$_3$Bi and $\alpha_{eff} \approx 2$ for the Cd$_3$As$_2$. So, the semimetals Na$_3$Bi and Cd$_3$As$_2$ belong  
to the class of strongly coupled systems. The study of such systems is very complicated problem and today there is no analytical 
approach that can satisfactorily deal with it. For this reason in this paper we are going to 
apply lattice simulation of Dirac semimetals which exactly regards many body effects in strongly coupled systems. 

Similarly to the gauge fields one can integrate out fermion fields from the partition function (\ref{eq:continuousaction}), 
what leads to the following expression
\begin{equation}\begin{split}
	Z=\int DA_4~\exp{\bigl ( -S_g \bigr )}\cdot \bigl ( {\det} D[A_4] \bigr )^{N_f},
\label{Zeff}
\end{split}\end{equation}

Due to the $\gamma_5$-hermiticity: $\gamma_5 D^{\dag}[A_4]\gamma_5=D[A_4]$, the eigenvalues of the Dirac 
operator go in pairs $\pm i \lambda$. This allows us to state that the ${\det} D[A_4]$ is positively defined.  
The product 
\beq
P[A_4] = \exp{\bigl ( -S_g \bigr )}\cdot \bigl ( {\det} D[A_4] \bigr )^{N_f}
\eeq
can be interpreted as a probability measure and Monte Carlo methods can be used to calculate partition function.  

The chiral symmetry of the action (\ref{eq:continuousaction}) is $SU_R(N_f)\times SU_L(N_f)$\footnote{Notice that 
there is also symmetry $U_V(1)\times U_A(1)$ of the action (\ref{eq:continuousaction}).}.
It is known that sufficiently large interaction strength dynamically breaks this symmetry according to 
the pattern $SU_R(N_f)\times SU_L(N_f) \to SU(N_f)$\cite{Scherer:2002tk,Pich:1998xt,Bijnens:1995ww,Klevansky:1992qe}. This leads to the formation of 
excitonic condensate $\bar \psi \psi$ and appearance of dynamical fermionic mass (the mass gap).
The system with dynamically generated mass gap is an insulator.  

In this paper we are going to study the semimetal/insulator phase transition with dynamical chiral symmetry breaking
within lattice simulation of Dirac semimetals. 
To do this we add the mass term to the action (\ref{eq:continuousaction})
\beq
S_m = \sum\limits_{a=1}^{N_f}\int d^3x dt~m \bar{\psi}_a \psi_a.
\eeq
Notice that this term violates $SU_R(N_f)\times SU_L(N_f)$ but conserves $SU(N_f)$, thus playing a 
role of the seed of the insulator phase. Monte Carlo lattice simulation of Dirac semimetals 
will be carried out with finite mass $m$ which will be extrapolated to the chiral limit at the end of the calculation.  
If in the limit $m \to 0$ the chiral condensate is nonzero then the system is in the insulator phase.
Otherwise the system is in the semimetal phase.

The order parameter of the semimetal/insulator phase transition with dynamical chiral symmetry breaking in the EFT is the chiral condensate 
$\sigma= \langle \bar{\psi}\psi \rangle$. If one introduces explicitly non-zero mass term to the action 
$S_E\to S_E + m\int d^3xdt\bar{\psi_a}\psi_a$, then the value of the chiral condensate is given by:

\beq
\sigma=\frac{T} {V}\frac{\partial \ln Z}{\partial m}=\frac{T}{V} \langle\tr D^{-1}[A_4]\rangle
\eeq

\section{Lattice field theory for Dirac semimetals}

\subsection{Staggered fermions without interactions}

In this section we are going to formulate low energy effective field theory of Dirac semimetals (\ref{eq:continuousactionscaled})
on discrete lattice. It is known that naive discretization of the continuum fermion action (\ref{sf}) leads 
to appearance of 16 fermion copies per one lattice fermion. This is well known the fermion doubling problem \cite{Montvay:1994cy}.  
To get rid of this problem we are going to use Kogut-Susskind staggered fermions \cite{Susskind:1976jm}. 

The low energy effective action for the staggered fermions (see below) is 
reduced to the action with four degenerate Dirac fermions. The quantum number which enumerates 
these degenerate fermions is called taste. Low energy theory of the observed 
Dirac semimetals contains two Dirac fermions. To get low energy 
effective action for two Dirac fermions from the action for the staggered fermions 
one can take square root from the fermion determinant. If the system 
has the taste symmetry which is the case for the low energy effective action of staggered fermions  
this procedure is well defined. However, below it will be seen that there are corrections 
to the low energy effective action of staggered fermions which violate taste symmetry of the fermion action. 
So, the procedure of taking square root of the staggered fermion determinant 
is well defined only in the continuum limit. We believe that this also leads
to noticeable volume dependence of finite results which were observed in \cite{Braguta:2016vhm}. 

To formulate the theory of the observed Dirac semimetals based on the staggered 
fermions we are going to use lattice with different spatial and temporal lattice 
spacings. 
Below it will be shown that taking continuum limit in temporal
direction at arbitrary fixed spatial lattice spacing the procedure of square rooting 
of the fermion determinant is well defined. The low energy effective 
action obtained in this case corresponds to two Dirac fermions and finite volume effects are small. 
Now let us proceed to the details of this approach.

To write the lattice fermion action for staggered fermions let us introduce a regular cubic lattice
in four-dimensional space with spatial lattice spacing $a_s$ and temporal lattice spacing $a_t$. 
The number of lattice sites is $L_s$ in each spatial direction and $L_t$ in temporal direction.  
The sites of the lattice have coordinates $x_{\mu=1,2,3}=0,1,..., L_s-1,~x_{4}=0,1,..., L_t-1$. 
With this notations the Euclidean action for staggered fermions can be written as 
\beq
S_f=\sum_x a_s^3 \biggl \{  m  \bar \psi_x \psi_x + \nonumber \\  + \frac 1 2 \sum_{\mu=1}^4 \xi_{\mu} \eta_{\mu}(x) \bigl [ 
(\bar \psi_{x} \psi_{x+\mu}) - (\bar \psi_{x+\mu} \psi_{x})
\bigr ] \biggr \}, 
\label{sf_lattice}
\eeq
where $\bar \psi_x, \psi_x$ are one component Grassmann fields and the sum is taken over all sites of the lattice,
$\eta_{\mu}(x)=(-1)^{\sum_{i<\mu} x_i}$, the $m$ is the mass in lattice units, and
\begin{equation}\begin{split}
	\xi_{\mu=1,2,3} & = \xi = a_t/a_s\\
	\xi_{\mu=4} & = 1.
\label{eq:xidefinition}
\end{split}\end{equation}
Notice that action (\ref{sf_lattice}) is the action for free fermions.
Interactions between 
fermions will be introduced below. In order to reduce the size of the 
formulas we will take $a_s=1$, restoring explicit spatial lattice 
spacing when necessary. 

Action (\ref{sf_lattice}) can be considered as some tight-binding model on four-dimensional lattice. 
Let us determine low energy spectrum of this theory\footnote{Here we follow \cite{Montvay:1994cy}.}.  
To do this we assume that the lattice sizes are even: $L_s=2L_s'$, $L_t=2L_t'$ and  
divide our lattice to hypercube blocks of size $2^4$. The coordinates of the original lattice 
can be written as 
\beq
x_{\mu}=2 y_{\mu} + \eta_{\mu},~~ \mu=1,2,3,4.
\eeq 
So, the coordinates $y_{\mu}=0,...,L'_{\mu}-1$ label hypercube blocks and the coordinates 
$\eta_{\mu}=0,1$ label the sites within one hypercube block.  

To proceed further we introduce the fields for four tastes of Dirac fermions: $\bar q^{a \alpha}, q^{\alpha a}$
where the $\alpha=1,2,3,4$ is the Dirac index and the $a=1,2,3,4$ is the taste index. Now one can build 
the fields $\bar q^{\alpha a}, q^{\alpha a}$ as a linear combinations of the 
fields $\bar \psi_x, \psi_x$ living in one hypercube block 
\beq
\bar q_y^{a \alpha} = \frac 1 8 \sum_{\eta} \Gamma^+_{\eta, a \alpha} \bar \psi_{2 y + \eta},~
q^{\alpha a} = \frac 1 8 \sum_{\eta} \Gamma_{\eta,\alpha a} \psi_{2 y + \eta},
\eeq
where the matrix $\Gamma_{\eta}=\gamma_1^{\eta_1} \gamma_2^{\eta_2} \gamma_3^{\eta_3} \gamma_4^{\eta_4}$
and $\gamma_{\mu}$ are the Dirac matrices. 

With these notations the fermion action can be written in the following form \cite{Montvay:1994cy}
\beq
S_f&=& \bar q \hat D q = 2^4\sum_y  \biggl \{ 
m~ \bar q_y ( {\bf 1} \otimes {\bf 1} ) q_y + \nonumber \\ 
&+& \sum_{\mu=1,2,3} \xi~ \bar q_y \bigr [  ({ \gamma_{\mu}} \otimes {\bf 1}) \Delta_{\mu} - 
({ \gamma_5}\otimes{ \gamma_5 \gamma_{\mu}^T}) \delta_{\mu}  \bigl ] q_y + \nonumber \\ 
&+&\bar q_y \bigr [ a_t ({ \gamma_{4}} \otimes {\bf 1}) \Delta_{4} - a_t^2 
({ \gamma_5}\otimes{ \gamma_5 \gamma_{4}^T}) \delta_{4}  \bigl ] q_y
\biggr \}.
\label{saction}
\eeq
In the last equation we used the following designations. $A \otimes B$ means 
that the matrix $A$ acts on the Dirac indices $\alpha$ and 
the matrix $B$ acts on the taste indices $a$ of the fermion fields. 
For further use we also introduced the Dirac matrix $\hat D$ 
which acts in the spaces of the coordinates, Dirac indices and taste indices.  
The operators $\Delta_{\mu}$ and $\delta_{\mu}$ have the form
\beq
\Delta_{\mu} f &=& \frac 1 {4 a_{\mu}} \bigl ( f_{y+\mu} - f_{y-\mu} \bigr ) \nonumber \\ 
\delta_{\mu} f &=& \frac 1 {4 a^2_{\mu}} \bigl ( f_{y+\mu} + f_{y-\mu} -2 f_y \bigr ).
\eeq 
In continuum limit the operators $\Delta_{\mu}$ and $\delta_{\mu}$ are reduced 
to first and second derivative correspondingly: $\Delta_{\mu} f \to \partial_{\mu} f$, 
$\delta_{\mu} f \to \partial^2_{\mu} f$

Low energy effective action is given by the terms with the smallest order 
of spatial and temporal derivatives. Thus it is seen from formula (\ref{saction}) 
that low energy effective action is the sum of 
the mass term and the terms with the operators $\Delta_{\mu=1,2,3,4}$. 
So, low energy effective action for the staggered fermions is reduced 
to the action with four degenerate tastes of Dirac fermions.

In the observed Dirac semimetals low energy effective action 
contains two Dirac fermions instead of four in the staggered fermions. 
However, if the action is degenerate in tastes one can take square root
of the fermion determinant $\det \hat D$ and thus obtain two Dirac fermions 
from four tastes of staggered fermions. 

In addition to the terms degenerate in taste index
in formula (\ref{saction}) one can see the terms which are proportional to 
the operator of the second derivative $\delta_{\mu}$. Although these terms 
only give corrections to the leading order effective theory, they 
are accompanied by the matrices $\gamma_5 \gamma_{\mu}^T$ in the taste space.
These matrices violate taste degeneracy in action (\ref{saction}). 
For this reason the procedure of taking square root of the staggered fermion determinant 
is well defined only in the continuum limit.

In order to get rid of the problem with rooting we are going 
to use an asymmetric lattice $a_s \neq a_t$. In particular, 
on the lattice with finite size of $a_s$ we will take limit 
$a_t \to 0$ or equivalently $\xi=a_t/a_s \to 0$. 

To show that this trick helps  to solve the problem with 
rooting, let us concentrate on equation (\ref{saction}).
In the limit $a_t \to 0$ the term which is proportional to the 
operator $\delta_4$ can be omitted and action (\ref{saction}) 
becomes
\beq
S_f&=& \sum_y a_s^3 \biggl \{ 
m ~ \bar q_y ( {\bf 1} \otimes {\bf 1} ) q_y + \nonumber \\ 
&+& \sum_{\mu=1,2,3} \xi~ \bar q_y \bigr [  ({ \gamma_{\mu}} \otimes {\bf 1}) \Delta_{\mu} - 
({ \gamma_5}\otimes{ \gamma_5 \gamma_{\mu}^T}) \delta_{\mu}  \bigl ] q_y + \nonumber \\ 
&+&\bar q_y \bigr [ a_t ({ \gamma_{4}} \otimes {\bf 1}) \Delta_{4}  \bigl ] q_y
\biggr \}.
\eeq  
To proceed let us recall that the matrices $\gamma_5 \gamma^T_{\mu=1,2,3}$ have the form
\beq
\gamma_5 \gamma_{\mu=1,2,3}= \begin{pmatrix} i\sigma_{\mu}^{T} & 0 \\ 0 & -i\sigma_{\mu}^{T}\end{pmatrix}.
\eeq
Thus the fermion determinant in the taste space can be written in the following form
\beq
&&\det \hat D|_{a_t \to 0}=\nonumber\\
&&\quad=\begin{pmatrix} A + \sum_{\mu=1,2,3} B_{\mu} i\sigma_{\mu}^{T} & 0 \\ 0 & A - \sum_{\mu=1,2,3} B_{\mu} i\sigma_{\mu}^{T} \end{pmatrix} =
\nonumber \\ 
&&\quad= \det \biggl ( A + \sum_{\mu=1,2,3} B_{\mu} i\sigma_{\mu}^{T} \biggr ) \times \nonumber \\
&& \hspace{1.2in}\times\det \biggl ( A - \sum_{\mu=1,2,3} B_{\mu} i\sigma_{\mu}^{T} \biggr ) =
\nonumber \\
&& \quad = \biggl ( \det \bigl ( AA^+ + \sum_{\mu=1,2,3} B_{\mu}^2 \bigr ) \biggr )^2,
\label{det}
\eeq
where $A$ and $B$ are the matrices acting in the coordinate and Dirac index spaces
\beq
A &=&\biggl ( m~{\bf 1} + \xi \sum_{\mu=1,2,3}  { \gamma_{\mu}}\Delta_{\mu} + \gamma_{4}\Delta_{4} \biggr ), \nonumber \\
B_{\mu} &=&-({ \gamma_5}) \delta_{\mu}.
\eeq 
Formula (\ref{det}) is derived in Appendix.

Notice that the matrix $A$ represents Dirac operator for one Dirac fermion and 
the matrices $B_{\mu}$ parametrize corrections to the low energy effective action. 
From equation (\ref{det}) one sees that the fermion determinant of staggered fermions in the limit
$a_t \to 0$ represents two copies of the theory with the determinant $AA^+ + \sum_{\mu=1,2,3} B_{\mu}^2$.
Thus the square rooting of the staggered fermion determinant in the limit $a_t \to 0$ 
leaves us with one copy of the determinant $\det(AA^+ + \sum_{\mu=1,2,3} B_{\mu}^2)$. The terms 
$\sim B_{\mu}^2$ give only corrections to the low energy effective action which is 
determined by the operator $A$. Thus we have shown that in the limit $a_t \to 0$
the rooting procedure gives the low energy effective theory with two Dirac fermions.

\subsection{Interacting staggered fermions}

First let us consider the action for the electromagnetic field. 
For discretization of the electromagnetic field we use  noncompact action:
\begin{equation}
S_g=\frac{\tilde{\beta}}2\sum_{x,i}(\theta_4(x)-\theta_4(x+i))^2.
\label{eq:noncompactqed}
\end{equation}
If one replaces $\theta_4$ with $a_tA_4$ and $\tilde{\beta}$ with
\begin{equation}\begin{split}
\tilde{\beta}=\beta \frac {a_s} {a_t}=\frac{v_{\perp}}{4\pi \alpha \xi},\\
\beta=\frac{1}{4\pi \alpha_{eff}},
\label{eq:betadef}
\end{split}\end{equation}
then in the continuum limit one recovers gauge part of the action (\ref{eq:continuousactionscaled}). Here $\beta$ is the inverse effective coupling constant and $\tilde{\beta}$ also includes the ratio of time- and space-like lattice steps $\xi$. In the limit of zero temporal lattice spacing $a_t\to0$ $\tilde{\beta}\to\infty$ and $\beta$ is kept constant.

Note that the action (\ref{eq:noncompactqed}) is gauge invariant under the gauge transformations which are described by arbitrary function $g(x)$, where $x$ is lattice coordinate:
\beq
	\theta_4(x)\to\theta_4(x)+g(x+\hat{4})-g(x)
\label{gauge_tr}
\eeq

In previous section we considered free fermions. It causes no difficulties to 
write fermion action for interacting staggered fermions
\begin{equation}
\begin{split}
S_f=\bar{\psi}_xD_{x,y}\psi_y=\sum\limits_x\left(m\bar{\psi}_x\psi_x+\right.\\
\left.
+\frac12 \eta_4(x) [\bar{\psi}_x e^{i\theta_4(x)}\psi_{x+\hat{4}}-\bar{\psi}_{x+4}e^{-i\theta_4(x)}\psi_x]\right.+\\
+\frac12\sum\limits_{i=1}^3\xi_i \eta_i(x) [\bar{\psi}_x\psi_{x+\hat{\i}}-\bar{\psi}_{x+\i}\psi_x]
\left.\right),
\label{eq:staggered}
\end{split}
\end{equation}
where $\theta_4(x)$ is the gauge field and $\xi_i$ is anisotropy factor.
If the Fermi velocity is isotropic in spatial directions, 
then $\xi_i=\xi=a_t/a_s$. If there is an anisotropy in the Fermi velocity
$v_{\perp} \neq v_{\parallel}$ then $\xi_i=\xi=a_t/a_s, i=1,2$ and $\xi_3=\xi v_{\parallel}/v_{\perp} =(a_t v_{\parallel})/(a_s v_{\perp})$. 
Notice that under gauge transformation (\ref{gauge_tr}) the fermion fields $\bar \psi_x, \psi_x$ transform as 
\beq
\psi_x \to e^{i \theta(x)} \psi_x,~~ \bar \psi_x \to \bar \psi_x e^{-i \theta(x)}. 
\eeq
The requirement of gauge invariance of fermion action unambiguously determines the form of action (\ref{eq:staggered}). 

Integration over fermionic degrees of freedom is performed explicitly and one gets the following effective action:
\beq
\nonumber
Z=\int D\theta_4(x) \exp{\bigl ( -S_{eff} \bigr )}, \\
S^{(eff)}=-\ln\det D[\theta]+S_{g}.
\label{eq:effaction_4}
\eeq

This action corresponds to four degenerate fermion flavors\cite{Montvay:1994cy} instead of $N_f=2$ observed in Na$_3$Bi and Cd$_3$As$_2$. The reduction $N_f=4\to2$ is performed by the standard rooting procedure. The effective action used in the simulation is 
\beq
S^{(eff)}=-\frac{1}{2}\ln\det D[\theta]+S_{g}.
\label{eq:effaction}
\eeq
The generation of the electromagnetic fields $\theta_4(x)$ was performed by means of the standard Hybrid Monte-Carlo Method\cite{Montvay:1994cy}.

It should be noted that $a_s$ plays a role of ultraviolet cutoff in the studied effective field theory. Although the exact value of $a_s$ is not known, one expects that it should be of order of the distance between atoms in the crystal structure of the material\cite{Braguta:2013rna}. 
Lattice simulation with action (\ref{eq:effaction}) will be performed at few values of the anisotropy coefficient $\xi$ and 
then the limit $\xi \to 0$ will be taken.

\begin{figure*}[t]
\begin{tabular}{cc}
\includegraphics[scale=0.45,clip=false]{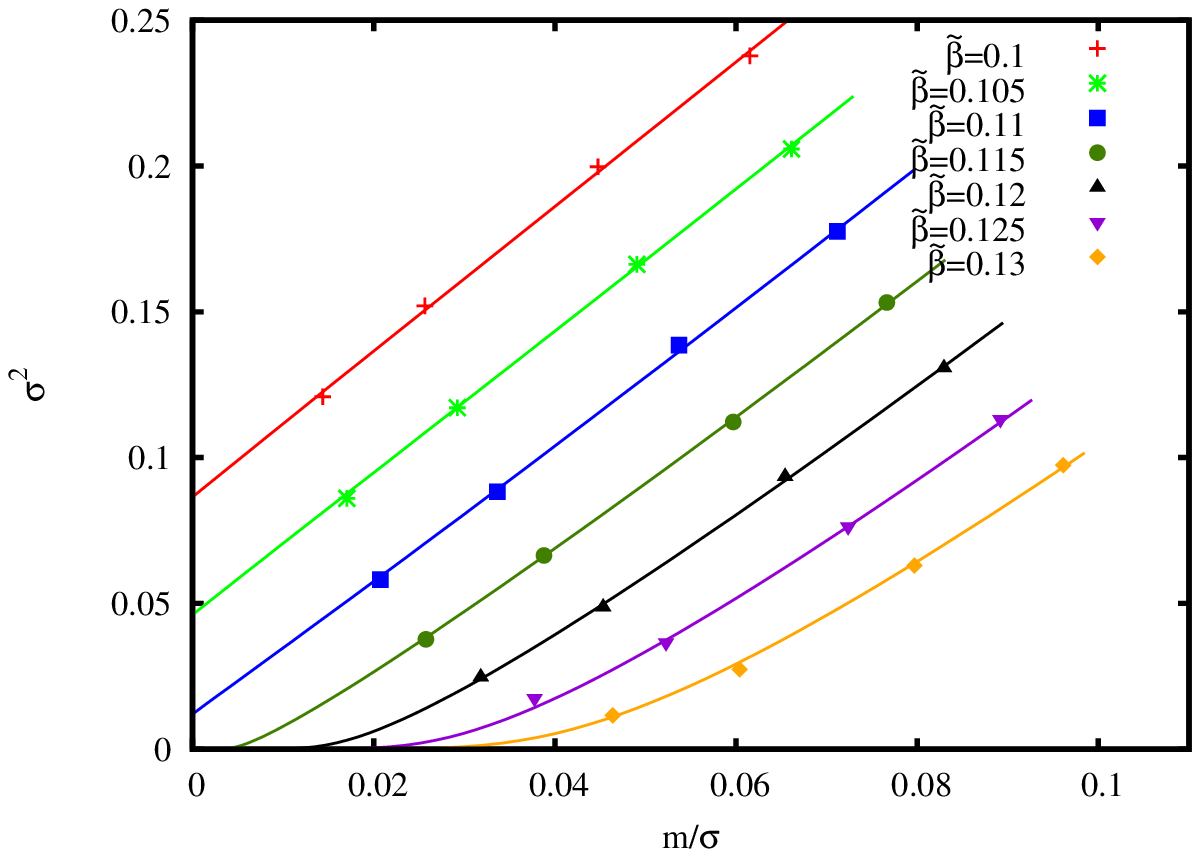} & 
\includegraphics[scale=0.45,clip=false]{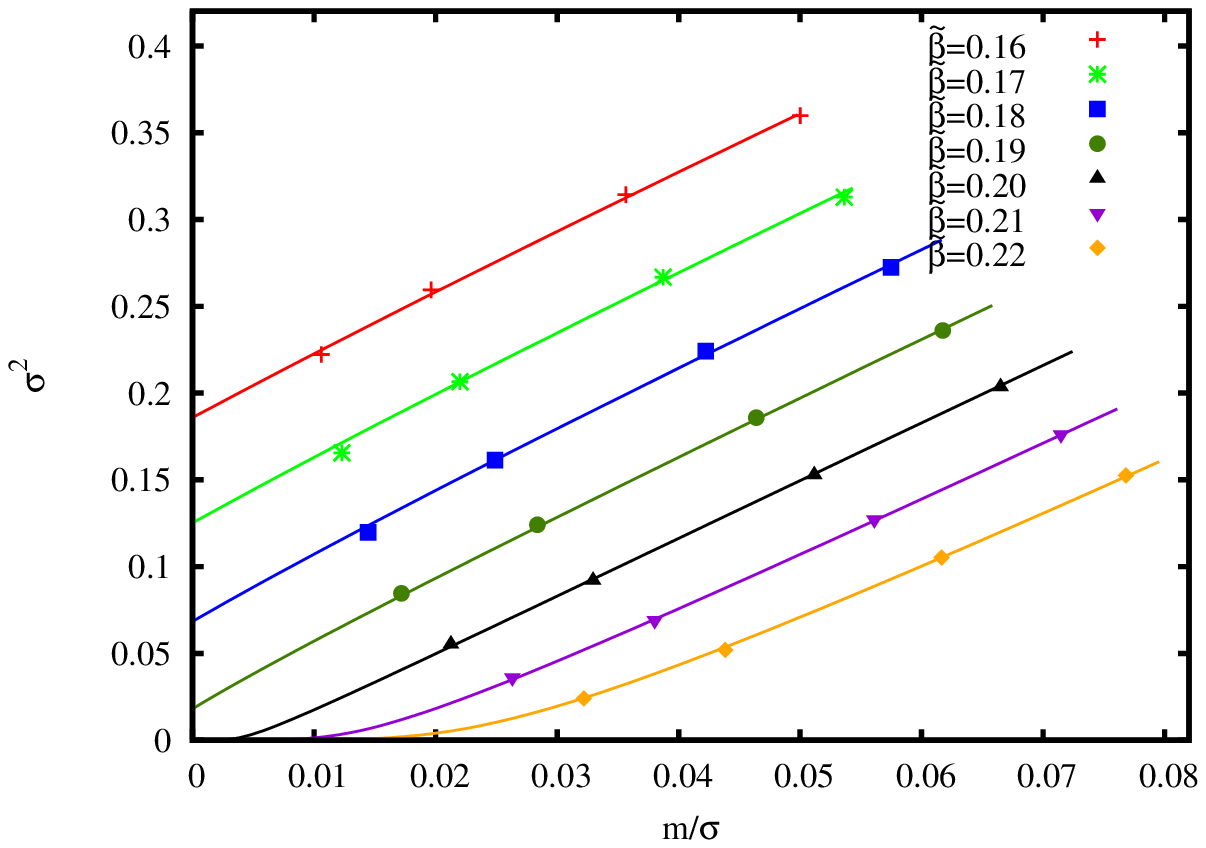}\\
$\xi=1/2$ & $\xi=1/3$ \\
\\

\includegraphics[scale=0.45,clip=false]{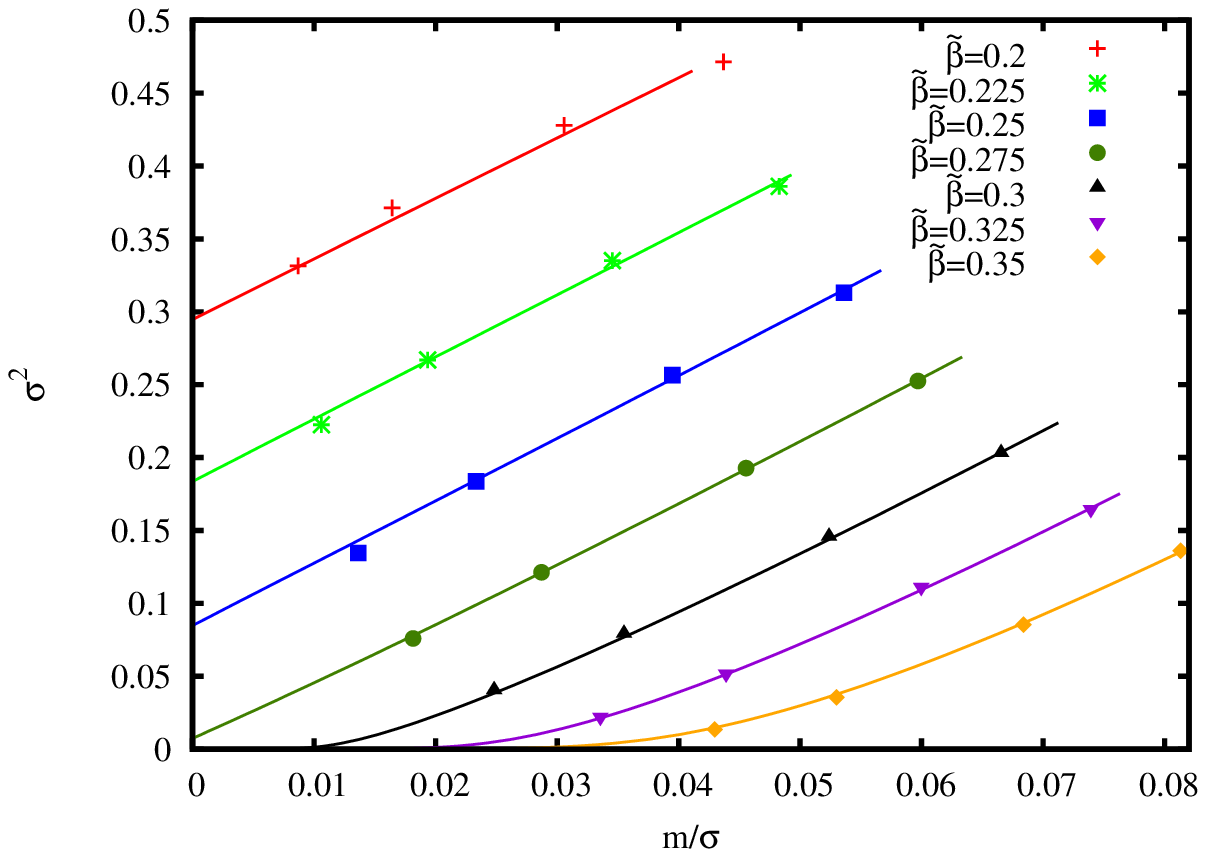} &
\includegraphics[scale=0.45,clip=false]{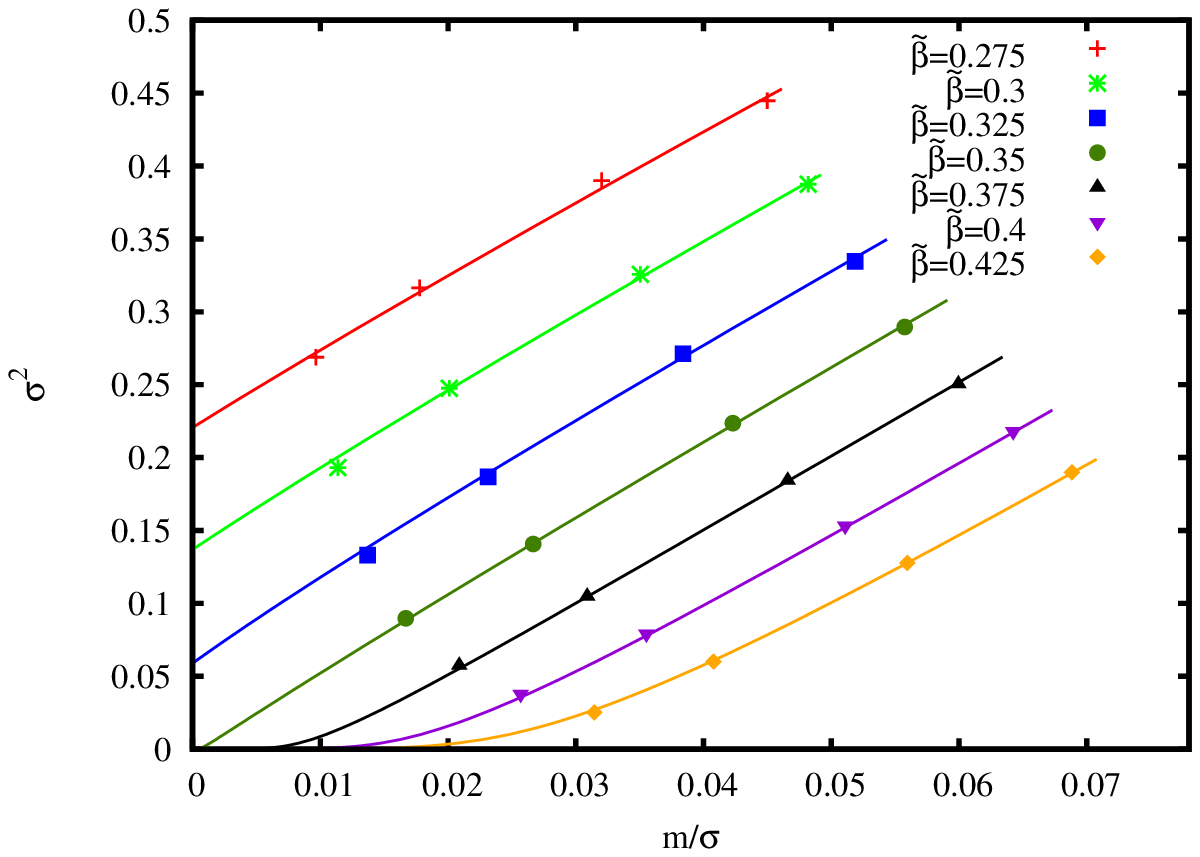}  \\
$\xi=1/4$ & $\xi=1/5$
\end{tabular}
\caption{Fisher plot for the $\sigma^2$ with respect to the ratio $m/\sigma$. The lines correspond to the fit by EoS (\ref{eq:eos}). Different plots correspond to different values of $\xi=\frac{a_t}{a_s}$ (Eq. (\ref{eq:xidefinition})).
}
\label{fig:fisher}
\end{figure*}

\begin{figure*}[t]
\begin{center}
\includegraphics[scale=0.60,clip=false]{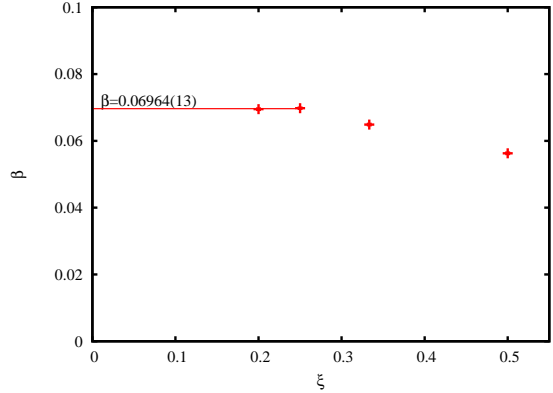}
\end{center}
\caption{The critical $\beta_c=\tilde{\beta}_c\xi=\frac{1}{4\pi \alpha_{eff}}$ as a function of $\xi=\frac{a_t}{a_s}$. Correct effective theory is restored in the limit $\xi\to0$. Red line corresponds to the average value for two points with the smallest $\xi=\frac14$ and $\xi=\frac15$.
}
\label{fig:res}
\end{figure*}

\section{Numerical results} 

In this section we are going to apply lattice field theory developed above to 
study the semimetal/insulator phase transition in the Dirac semimetals driven by strong Coulomb interaction. The order parameter for this phase transition is the chiral condensate $\sigma$, which is measured directly in our simulations. Its behaviour in the chiral limit $m\to0$ characterizes the phase of the system. In the semimetal phase $\sigma(m\to0)\to0$, while in the insulator phase $\sigma(m\to0)\ne0$. 

To measure the chiral condensate on the lattice we used stochastic estimator method:

\beq
\sigma=\frac1{VT} \langle\tr D^{-1}\rangle=\nonumber\\
\frac1{VT}\frac{1}{N_{conf}}\sum_{conf}\sum_{xy}\bar{\eta}_x(D^{-1}[\theta_4])_{xy}\eta_y,
\label{eq:stochasticestimator}
\eeq
where $\eta_x$ are Gaussian distributed random variables: $\langle\eta_x\rangle=0$, $\langle\bar{\eta}_x\eta_y\rangle=\delta_{x,y}$. Note also that in formula (\ref{eq:stochasticestimator}) we averaged over configurations of gauge fields $\theta_4$. For each set of the parameters we generated $\sim O(200)$ configurations. For the inversion of the Dirac operator $\sum_y(D^{-1}[\theta_4])_{xy}\eta_y$ we used Conjugate Gradient Method.

To check finite volume effects we performed numerical simulations with different lattice sizes $L_s^3\times L_t$ (see Tab.~\ref{tab:res} and ~\ref{tab:aniso}). We have found, that for all values of lattice size under study, the measured values of the chiral condensate as well as the extracted value of $\beta_c$ coincide with good accuracy, thus we expect that finite volume effects are small. For our final predictions we took the results for $L_s=16$, while $L_t$ was different for different set of parameters.

\subsection{Isotropic Fermi velocity}
\label{ch:iso}
	First we present the results for the case of isotropic Fermi velocity $\xi_1=\xi_2=\xi_3=\xi$. In order to study the limit $a_t\to0$ we performed simulations for four values of $\xi=1/2,1/3,1/4,1/5$. To control finite size effects simulations for the smallest and the largest value of $\xi=1/2,1/5$ were performed at different lattice sizes $L_s^3\times L_t$. We have found, that for lattice sizes used in our simulations the measured values of the chiral condensate coincide within the errorbars. Based on this observation we expect that finite volume effects are very small and can be neglected. Values of lattice sizes used in our simulations are presented in Tab.~\ref{tab:res}.

\begin{table}[h!]
\centering
\begin{tabular}{|c|c|c|c|c|}
\hline
$\xi$ & $L_s$ & $L_t$ & $\tilde{\beta}_c$ & $\beta_c$ \\ 
\hline
1/2 & 12 & 12 & 0.1133(7) & 0.0566(4)\\
1/2 & 16 & 16 & 0.1127(5) & 0.0564(2)\\
1/2	& 16 & 32 & 0.1126(4) & 0.0563(2) \\
1/2 & 20 & 12 & 0.1106(3) & 0.0553(2)\\
1/2 & 20 & 24 & 0.1115(2) & 0.05573(14)\\
\hline
1/3 & 16 & 48 & 0.1946(6) & 0.0648(2) \\
\hline
1/4 & 16 & 64 & 0.2791(7) & 0.06977(16)\\
\hline
1/5 & 10 & 30 & 0.3438(27) & 0.0688(5) \\
1/5 & 14 & 48 & 0.3505(11) & 0.0701(2) \\
1/5 & 16 & 80 & 0.3474(9) & 0.0695(2) \\
1/5 & 20 & 30 & 0.3420(9) & 0.0684(2) \\
1/5 & 20 & 48 & 0.3468(9) & 0.0697(2)  \\
1/5 & 20 & 100 & 0.3487(5) & 0.06975(11)\\ 
\hline

\end{tabular}
\caption{The critical value of $\tilde{\beta}$ and $\beta$ (Eq. (\ref{eq:betadef})) for different values of $\xi=\frac{a_t}{a_s}$ (Eq. (\ref{eq:xidefinition})) and lattice sizes. 
}
\label{tab:res}
\end{table}

\begin{figure*}[t]
\begin{tabular}{ccc}
\includegraphics[scale=0.40,clip=false]{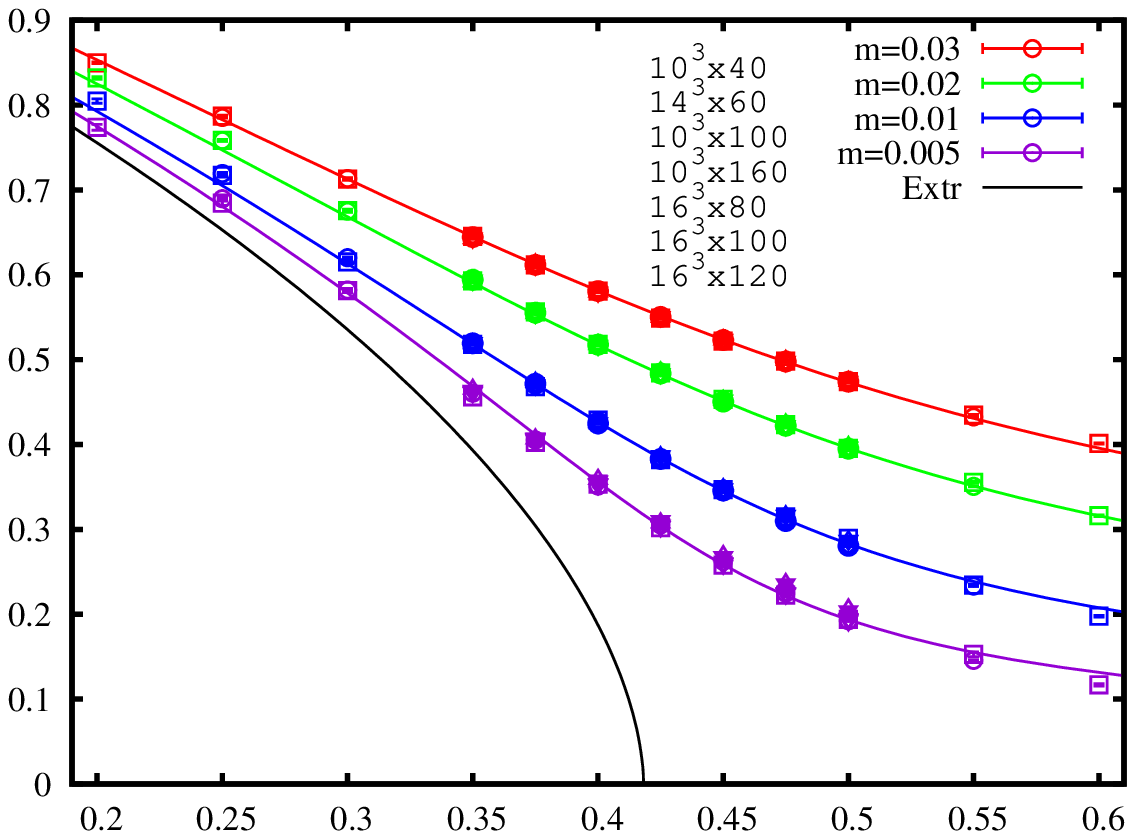} & 
\includegraphics[scale=0.40,clip=false]{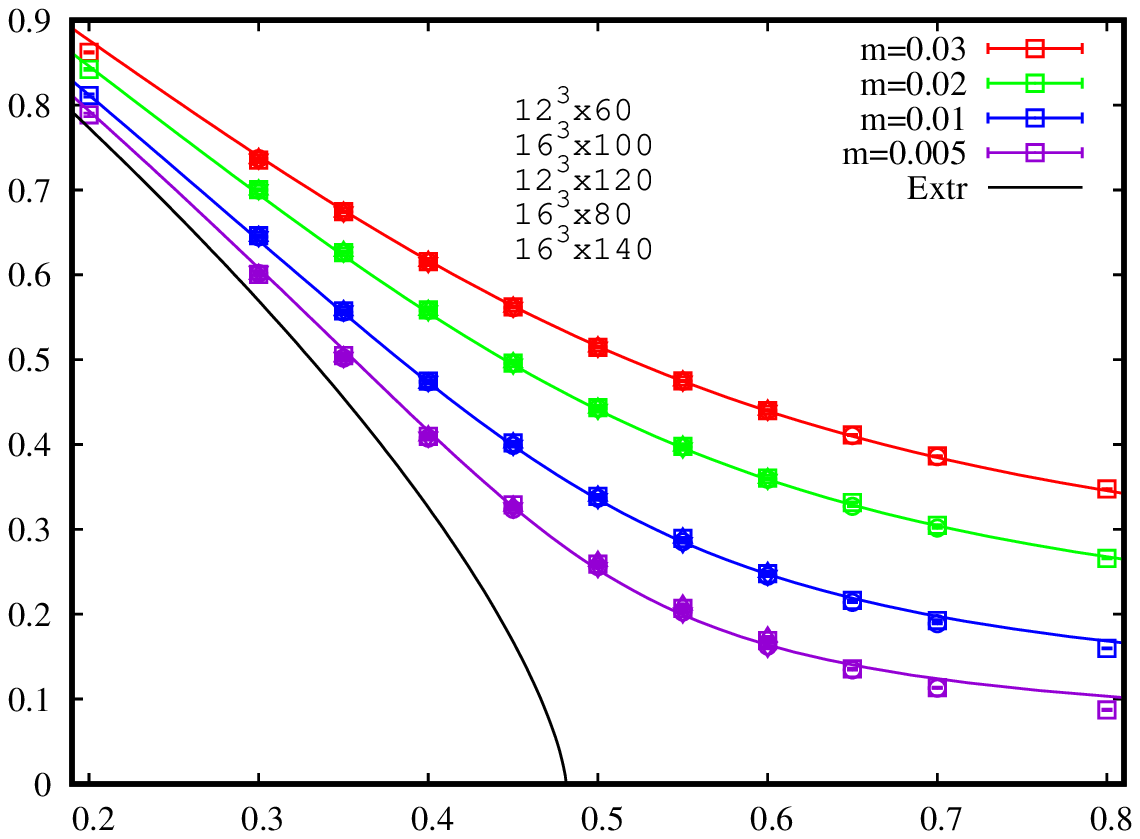} &
\includegraphics[scale=0.40,clip=false]{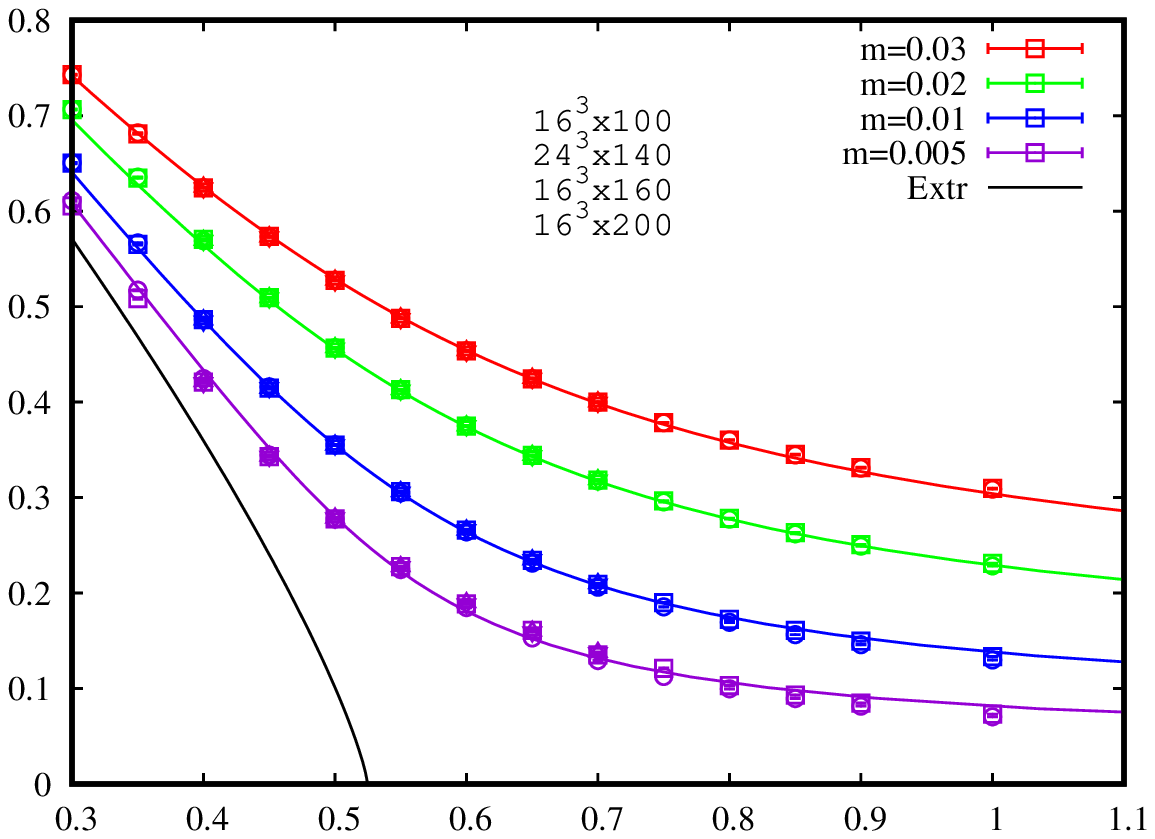} \\
\includegraphics[scale=0.40,clip=false]{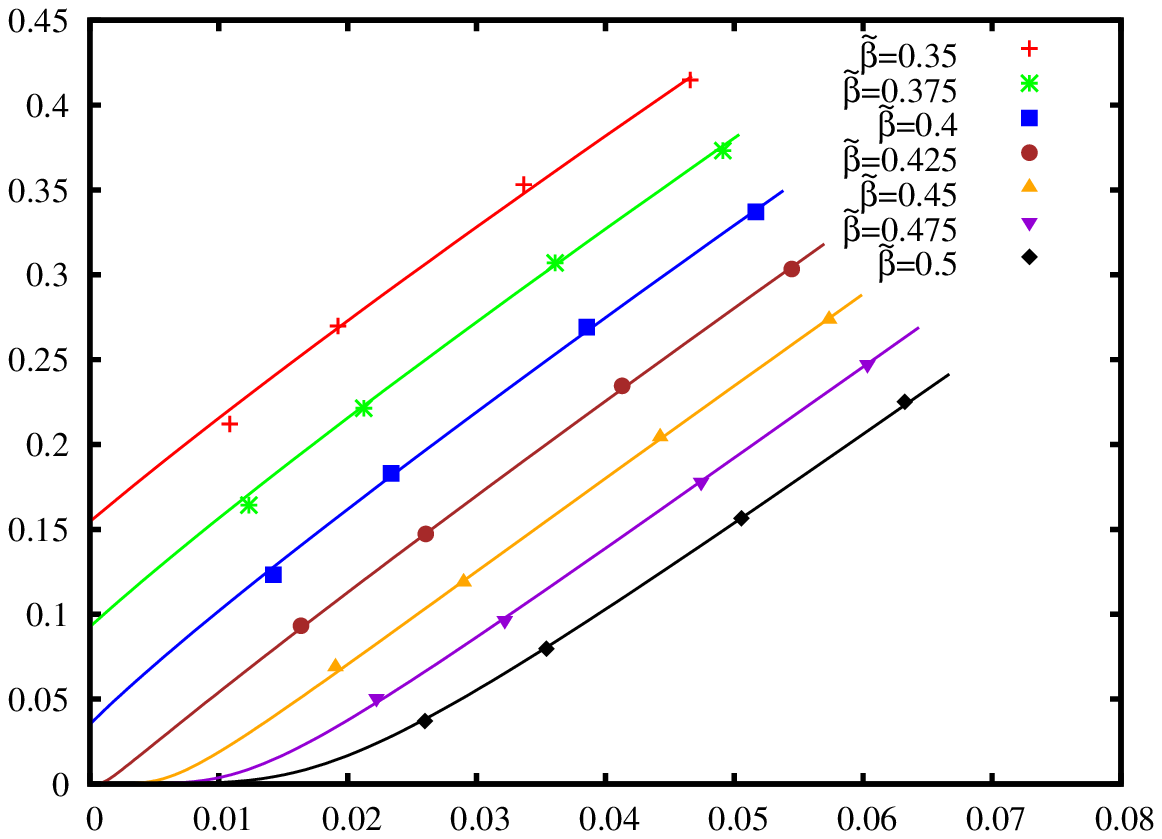} & 
\includegraphics[scale=0.40,clip=false]{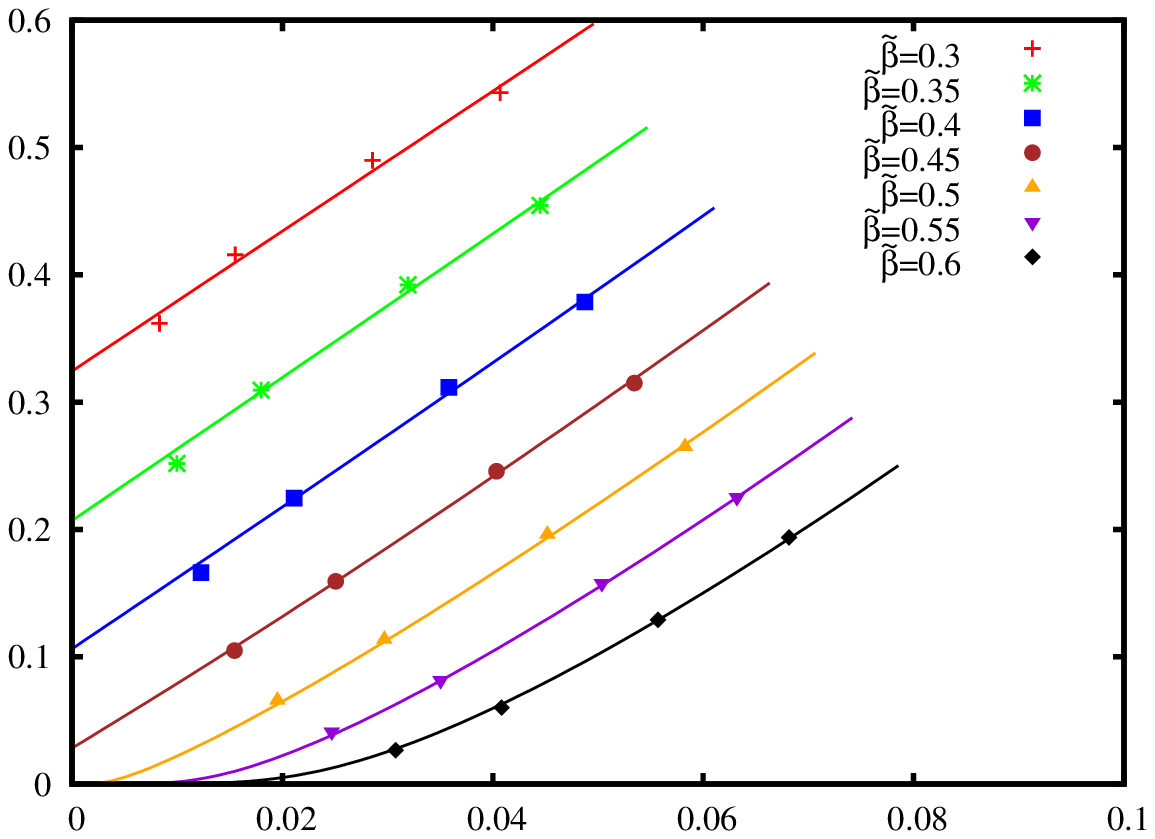} &
\includegraphics[scale=0.40,clip=false]{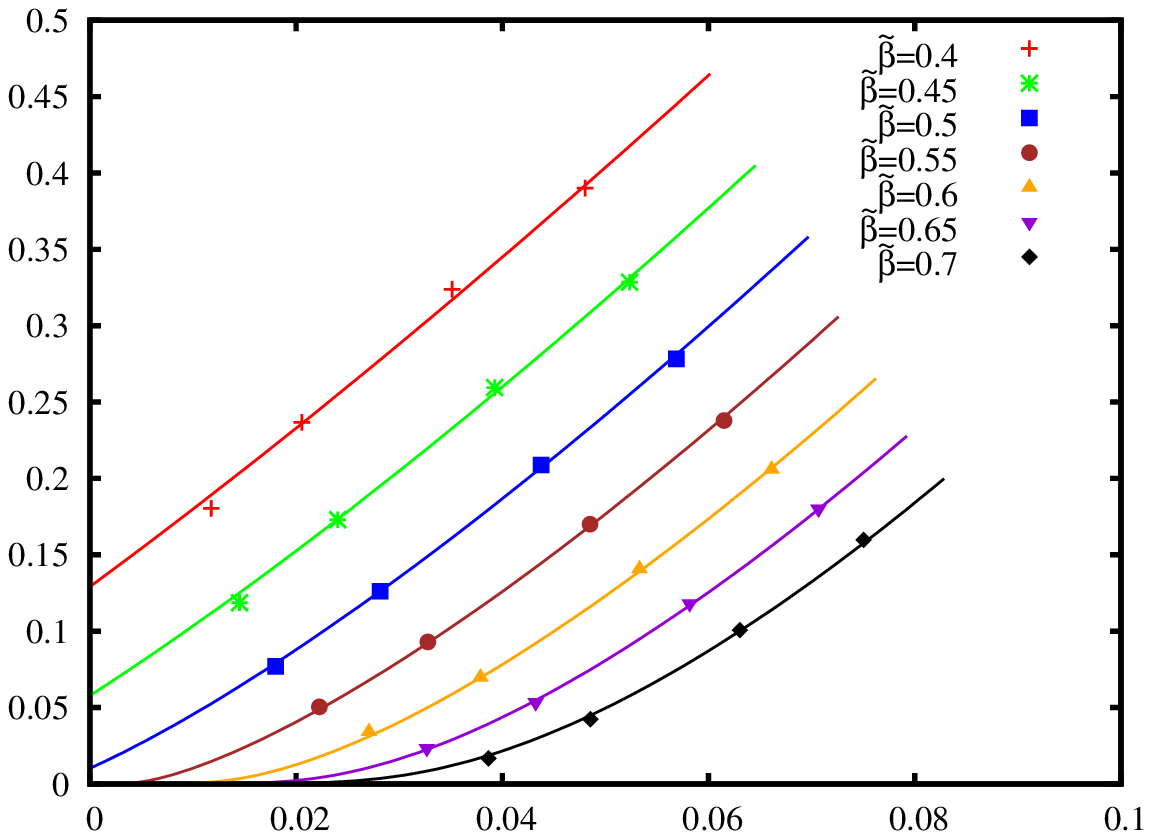} \\
$\zeta=1/2$ & $\zeta=1/4$ & $\zeta=1/10$
\end{tabular}
\caption{Chiral condensate as a function of the $\beta$(upper row) and Fisher plot(lower row) for the $\sigma^2$ with respect to the ratio $m/\sigma$ for different Fermi velocity anisotropies $\zeta=\frac{v_{\parallel}}{v_{\perp}}$. Black lines on the upper plots correspond to the $m\to0$ extrapolation with the help of EoS (\ref{eq:eos}). Coloured lines on the lower plots correspond to the fit by EoS (\ref{eq:eos}).
}
\label{fig:anisodata}
\end{figure*}

\begin{figure*}[t]
\begin{center}
\includegraphics[scale=0.60,clip=false]{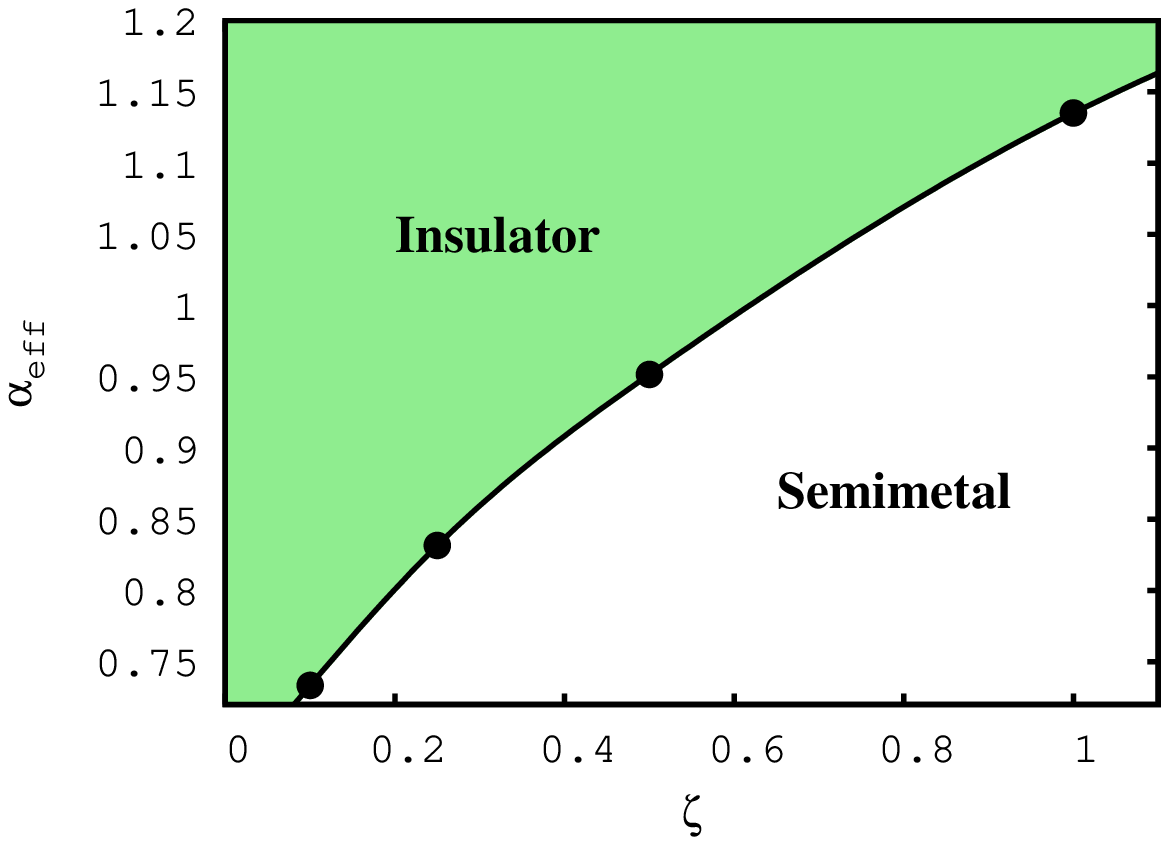}
\end{center}
\caption{Dependence of the critical effective coupling $\alpha_{eff}^c$ on Fermi velocity anisotropy $\zeta=\frac{v_{\parallel}}{v_{\perp}}$. Points correspond to the data obtained on the lattice. Lines are to guide the eyes.
}
\label{fig:phase}
\end{figure*}

The dependence of the chiral condensate $\sigma$ on $\tilde{\beta}$ for different values of mass $m$ is presented in the Fig.~\ref{fig:isopsipsi}. On the plots the data points for different lattice sizes almost coincide. All plots exhibit the similar behavior showing the formation of the chiral condensate and phase transition from semimetal to insulator phase. The critical value of $\tilde{\beta}$ can be roughly estimated to be $\tilde{\beta}_c(\xi=1/2)\sim0.12$, $\tilde{\beta}_c(\xi=1/3)\sim0.2$, $\tilde{\beta}_c(\xi=1/4)\sim0.3$, $\tilde{\beta}_c(\xi=1/5)\sim0.35$. 

For more precise determination of the critical coupling constant we used the Equation of State(EoS), which was successfully applied to strongly coupled QED \cite{Gockeler:1996me} and graphene\cite{Drut:2008rg}:
\begin{equation}
    m X(\tilde{\beta})=Y(\tilde{\beta})f_1(\sigma)+f_3(\sigma),
\label{eq:eos}
\end{equation}
where $X(\tilde{\beta})=X_0+X_1(1-\tilde{\beta}/\tilde{\beta}_c)$, $Y(\tilde{\beta})=Y_1(1-\tilde{\beta}/\tilde{\beta}_c)$ are expanded in the vicinity of critical $\tilde{\beta}_c$. For the functions $f_1$ and $f_3$ we allowed nonclassical critical exponents: $f_1(\sigma)=\sigma^b$, $f_3(\sigma)=\sigma^{\delta}$. Classical values of $b=1$ and $\delta=3$ can be easily visualized, if one plots $\sigma^2$ as a function of $m/\sigma$(Fisher plot). The resulting dependence $\sigma^2(m/\sigma)$ form straight lines, crossing the origin at critical value $\beta_c$. The deviations from straight lines might be attributed to non-classical values of $b$ and $\delta$. The Fisher plots for the considered values of $\xi$ are presented in Fig.~\ref{fig:fisher}. The lines on the plots correspond to the fit by EoS (\ref{eq:eos}). One sees small deviations from classical values $b=1$, $\beta=3$. The critical values of $\tilde{\beta}_c$, obtained with the help of this fit, are also presented in Tab.~\ref{tab:res}. One can easily obtain critical value of inverse effective coupling by the formula $\beta_c=\tilde{\beta}_c\xi$.

We plot the dependence of the critical value $\beta_c$ on $\xi$ in the Fig.~\ref{fig:res}. Note, that the limit $\xi\to0$ should be taken. One sees that for $\xi=1/4$ and $\xi=1/5$ the critical values $\beta_c$ are almost the same. As the final estimate we take half of the sum of $\beta_c$ obtained at $\xi=1/4$ (lattice size $16^3\times64$) and $\xi=1/5$ (lattice size $16^3\times80$). The obtained critical value is $\beta_c=0.06964(13)$, which corresponds to the critical coupling $\alpha_c=1.1414(21)$. Note that this critical coupling constant is smaller, 
than was obtained earlier in \cite{Braguta:2016vhm}. We believe that this discrepancy can be attributed to finite volume effects
which are under control in the present study.

\subsection{Fermi velocity anisotropy}

In the discovered materials Na$_3$Bi and Cd$_3$As$_2$ there is an anisotropy in the Fermi velocity $v_{\parallel}\ne v_{\perp}$. For this reason it is interesting to study how this anisotropy affects the phase diagram of the system.

The calculations presented in the previous section show that for $\xi=1/5$ we are close to the limit $a_t\to 0$. Accordingly, here we fix two values $\xi_1=\xi_2=\xi=1/5$ and vary $\xi_3=\zeta\xi$, choosing $\zeta\in(0;1)$. One can expect that in this case we are also close to the limit $a_t\to0$. Note that it corresponds to the relation $v_{\parallel}=\zeta v_{\perp}$ between Fermi velocity in different directions. 

For checking finite volume effects we also performed simulations at various lattice sizes, which are listed in the Tab.~\ref{tab:aniso}. 

In Fig.~\ref{fig:anisodata} the dependences of the chiral condensate on $\beta$ as well as Fisher plots are shown for three values of $\zeta=\frac12,\frac14,\frac1{10}$. By fitting data to the same EoS (\ref{eq:eos}), critical $\beta_c$ was extracted. Obtained values of critical $\beta_c$ are presented in Tab.~\ref{tab:aniso}. As in the isotropic case the presented results suggest that finite volume effects are small. Using these values the phase diagram of the system is drawn in Fig.~\ref{fig:phase}. For this plot we took spatial lattice size $L_s=16$ and the largest temporal lattice size $L_t=120$ for $\zeta=1/2$, $L_t=140$ for $\zeta=1/4$, $L_t=200$ for $\zeta=1/10$.

\begin{table}[h!]
\centering
\begin{tabular}{|c|c|c|c|c|}
\hline
$\zeta$ & $L_s$ & $L_t$ & $\tilde{\beta}_c$ & $\beta_c$\\
\hline
1/2 & 10 & 40 & 0.408(3) & 0.0816(6) \\
1/2 & 10 & 100 & 0.433(3) & 0.0865(6) \\
1/2 & 10 & 160 & 0.435(2) & 0.0869(5) \\
1/2 & 14 & 60 & 0.418(2) & 0.0836(4) \\
1/2 & 16 & 80 & 0.4159(12) & 0.0832(2)\\
1/2 & 16 & 100 & 0.4208(13) & 0.0842(3)\\
1/2 & 16 & 120 & 0.4181(11) & 0.0836(2)\\
\hline
1/4 & 12 & 60 & 0.476(3) & 0.0952(6)\\
1/4 & 12 & 120 & 0.493(3) & 0.0986(5)\\
1/4 & 16 & 80 & 0.4802(16) & 0.0960(3)\\
1/4 & 16 & 100 & 0.4812(14) & 0.0962(3)\\
1/4 & 16 & 140 & 0.4784(14) & 0.0957(3)\\
\hline
1/10 & 16 & 100 & 0.5349(22) & 0.1070(4)\\
1/10 & 16 & 160 & 0.5388(18) & 0.1078(4)\\
1/10 & 16 & 200 & 0.5425(19) & 0.1085(4)\\
1/10 & 24 & 140 & 0.5248(11) & 0.10496(22)\\
\hline
\end{tabular}
\caption{The critical value of $\tilde{\beta}_c$ and $\beta_c$ (Eq. (\ref{eq:betadef})) for different anisotropy of Fermi velocity $\zeta=\frac{v_{\parallel}}{v_{\perp}}$ and lattice sizes.}
\label{tab:aniso}
\end{table}

\section{Conclusions and discussion} 

This paper is devoted to the study of electronic properties of recently discovered Dirac semimetals Na$_3$Bi and Cd$_3$As$_2$.
We formulate the low energy effective theory for these materials, which is described by $N_f=2$ massless 3D Dirac fermions with
Coulomb interaction. We construct lattice discretized version of this theory and show that in the limit of
zero temporal lattice spacing $a_t\to0$ it reproduces low energy effective theory with the correct number of degrees of freedom.
Methods for numerical simulation of this lattice theory are briefly summarized.

In the second part of the paper we studied the phase diagram of Dirac semimetals using supercomputer simulations of the formulated 
lattice effective field theory. In particular, we concentrate on the semimetal-insulator phase transition driven by Coulomb interaction. 
Systematic uncertainties due to finite volume effects and nonzero temporal lattice spacing $a_t\ne0$ are under control. 
By means of numerical simulations we determine the value of the effective critical coupling and its dependence on the
Fermi velocity anisotropy. Based on these results we draw tentative phase diagram for the Dirac semimetals.

The phase diagram of the Dirac semimetals discussed in this paper was studied earlier in the strong coupling limit\cite{Sekine:2014yna,Araki:2015php}, 
by means of Dyson-Schwinger equations\cite{Gonzalez:2015iba}  and within ladder approximation\cite{Gonzalez:2015tsa}. It is interesting 
to discuss their results and to compare them with ours. In \cite{Araki:2015php} it was shown that in the limit of infinitely strong coupling 
$\beta=0$ for all values of Fermi velocity anisotropy $v_{\perp}/v_{\parallel}$ there is a gap, which is in line with our results. Note that the authors
of \cite{Araki:2015php} also discussed the importance of anisotropic lattices $a_t/a_s\to0$. The authors of \cite{Gonzalez:2015iba} studied the phase diagram 
of Dirac semimetals without Fermi velocity anisotropy within Dyson-Schwinger equations. The critical value of the coupling constant obtained by such 
approach is much larger $\alpha\approx14.7$. Later the same authors revised this result within ladder approximation \cite{Gonzalez:2015tsa}
and obtained the critical coupling to be equal $\alpha_{eff}^c=1.8660$ which is close to the value $\alpha_c=1.1414(21)$ obtained here. 

It is also worth mentioning the paper \cite{2017arXiv170203551S}, where the authors studied the graphene under uniaxial strain, 
leading to anisotropic Fermi velocity. The results, obtained in this paper are in qualitative agreement with ours: large Fermi velocity
anisotropy $v_{\perp}/v_{\parallel}\ll 1$ leads to smaller effective coupling constant.

Finally we would like to mention that the results obtained in this paper are approximately by a factor of 
$\sim 1.5$ smaller than that obtained in our previous paper \cite{Braguta:2016vhm} for all values of
the Fermi velocity anisotropy. We believe that the discrepancy can be attributed to 
the finite volume effects which are under control in the present paper. 

Despite this discrepancy our main conclusions remain the same. The first one is that 
the Fermi velocity anisotropy leads to decrease of critical coupling constant.
The second one is that the observed Dirac semimetals Na$_3$Bi and Cd$_3$As$_2$ 
lie deep in the insulator phase what contradicts to experiment. 
A possible resolution of this discrepancy is that in real world 
the interaction potential at small distances is screened by bound electrons
what can lead to the shift of the critical coupling to larger values. 
Another possible explanation is that due to the renormalization effects
strong interaction can considerably modify the basic parameters of the theory. 
Although the study of different
explanation of the raised question is very important it is
beyond the scope of this paper.

\section*{Acknowledgments}

The authors would like to thank Z.V.~Khaidukov for useful discussions. 
The work on numerical simulation and the determination of the critical coupling
constants at different Fermi velocity asymmetries,
and the formulation of lattice field theory on anisotropic lattice
were supported by the RSF grant under contract 16-12-10059. The work of M.I.K. was supported by Act 211 of the government of the
Russian Federation, Contract No. 02.A03.21.0006.
Numerical simulations were carried out on GPU cluster of NRC Kurchatov Institute and at MSU supercomputer "Lomonosov".

\appendix
\section{}
In Appendix we are going to derive formula (\ref{det}).
We start from expression
\beq
\det \biggl ( A + i \sum_{\mu=1,2,3} B_{\mu} \sigma_{\mu}^T \biggr ) = \nonumber \\ = 
\det \biggl ( m {\bf 1} + \xi \sum_{\mu=1,2,3}  { \gamma_{\mu}}\Delta_{\mu} + \gamma_{4}\Delta_{4} + i \gamma_5 \hat \delta  \biggr ),
\eeq
where the following designation is used 
\beq
\hat \delta &=& - \sum_{\mu=1,2,3} \delta_{\mu} \sigma_{\mu}^T
\eeq 
Notice that the Pauli matrices $\sigma_{\mu}$ and the operator $\hat \delta$ act in the space of taste indices. 

In the calculation we used Euclidean $\gamma$-matrices which are defined $\gamma_4=\gamma_0^M,~\gamma_\mu=i \gamma_\mu^M,~ \mu=1,2,3$,
where $\gamma^M_{\mu}$ are $\gamma$-matrices in Minkowski space. 
Expanding $\gamma$-matrices one can write
\beq
\det \biggl ( A + i \sum_{\mu=1,2,3} B_{\mu} \sigma_{\mu}^T \biggr ) = \nonumber \\ = \det  
\begin{pmatrix} m + \Delta_4 & i \xi \sum_{\mu=1,2,3} \tau_{\mu} \Delta_{\mu} + i \hat \delta \\ -i \xi \sum_{\mu=1,2,3} \tau_{\mu} \Delta_{\mu} +i \hat \delta & m - \Delta_4 \end{pmatrix}, \nonumber
\eeq
where the Pauli matrices $\tau_{\mu}$ act in the space of spinor indices.
Now it is evident that the diagonal operators $ m \pm \Delta_4$ (which are unity operators in spin and taste indices) 
commute with the non-diagonal ones. For this reason one can write 
\beq
\det \biggl ( A + i \sum_{\mu=1,2,3} B_{\mu} \sigma_{\mu}^T \biggr ) = \nonumber \\ = \det \biggl ( \bigl ( m^2-\Delta_4^2 - 
\xi^2 \sum_{\mu=1,2,3} \Delta_{\mu}^2 + \sum_{\mu=1,2,3} \delta_{\mu}^2 \bigr)\times{\bf1}\times{\bf1} \biggr ),
\nonumber
\eeq
where the unity matrices ${\bf 1}$ act in spinor and taste spaces. Evidently last equation can be rewritten as
\beq
\det \biggl ( A + i \sum_{\mu=1,2,3} B_{\mu} \sigma_{\mu}^T \biggr ) = \det \biggl ( AA^+ + \sum_{\mu=1,2,3} B_{\mu}^2 \biggr ).
\nonumber
\eeq
Similarly
\beq
\det \biggl ( A - i \sum_{\mu=1,2,3} B_{\mu} \sigma_{\mu}^T \biggr ) = \det \biggl ( AA^+ + \sum_{\mu=1,2,3} B_{\mu}^2 \biggr ).
\nonumber
\eeq

\bibliographystyle{unsrt}

\begin{thebibliography}{10}

\bibitem{Novoselov666}
K.~S. Novoselov, A.~K. Geim, S.~V. Morozov, D.~Jiang, Y.~Zhang, S.~V. Dubonos,
  I.~V. Grigorieva, and A.~A. Firsov.
\newblock Electric field effect in atomically thin carbon films.
\newblock {\em Science}, 306(5696):666--669, 2004.

\bibitem{Geim2007}
A.~K. Geim and K.~S. Novoselov.
\newblock The rise of graphene.
\newblock {\em Nat Mater}, 6(3):183--191, Mar 2007.

\bibitem{PhysRev.71.622}
P.~R. Wallace.
\newblock The band theory of graphite.
\newblock {\em Phys. Rev.}, 71:622--634, May 1947.

\bibitem{PhysRev.104.666}
J.~W. McClure.
\newblock Diamagnetism of graphite.
\newblock {\em Phys. Rev.}, 104:666--671, Nov 1956.

\bibitem{Semenoff:1984dq}
Gordon~W. Semenoff.
\newblock {Condensed Matter Simulation of a Three-dimensional Anomaly}.
\newblock {\em Phys. Rev. Lett.}, 53:2449, 1984.

\bibitem{Novoselov2005}
K.~S. Novoselov, A.~K. Geim, S.~V. Morozov, D.~Jiang, M.~I. Katsnelson, I.~V.
  Grigorieva, S.~V. Dubonos, and A.~A. Firsov.
\newblock Two-dimensional gas of massless dirac fermions in graphene.
\newblock {\em Nature}, 438(7065):197--200, Nov 2005.

\bibitem{Zhang2005}
Yuanbo Zhang, Yan-Wen Tan, Horst~L. Stormer, and Philip Kim.
\newblock Experimental observation of the quantum hall effect and berry's phase
  in graphene.
\newblock {\em Nature}, 438(7065):201--204, Nov 2005.

\bibitem{Liu864}
Z.~K. Liu, B.~Zhou, Y.~Zhang, Z.~J. Wang, H.~M. Weng, D.~Prabhakaran, S.-K. Mo,
  Z.~X. Shen, Z.~Fang, X.~Dai, Z.~Hussain, and Y.~L. Chen.
\newblock Discovery of a three-dimensional topological dirac semimetal, na3bi.
\newblock {\em Science}, 343(6173):864--867, 2014.

\bibitem{Neupane2014}
Madhab Neupane, Su-Yang Xu, Raman Sankar, Nasser Alidoust, Guang Bian, Chang
  Liu, Ilya Belopolski, Tay-Rong Chang, Horng-Tay Jeng, Hsin Lin, Arun Bansil,
  Fangcheng Chou, and M.~Zahid Hasan.
\newblock Observation of a three-dimensional topological dirac semimetal phase
  in high-mobility cd3as2.
\newblock {\em Nat Commun}, 5, May 2014.

\bibitem{PhysRevLett.113.027603}
Sergey Borisenko, Quinn Gibson, Danil Evtushinsky, Volodymyr Zabolotnyy, Bernd
  B\"uchner, and Robert~J. Cava.
\newblock Experimental realization of a three-dimensional dirac semimetal.
\newblock {\em Phys. Rev. Lett.}, 113:027603, Jul 2014.

\bibitem{Xu613}
Su-Yang Xu, Ilya Belopolski, Nasser Alidoust, Madhab Neupane, Guang Bian,
  Chenglong Zhang, Raman Sankar, Guoqing Chang, Zhujun Yuan, Chi-Cheng Lee,
  Shin-Ming Huang, Hao Zheng, Jie Ma, Daniel~S. Sanchez, BaoKai Wang, Arun
  Bansil, Fangcheng Chou, Pavel~P. Shibayev, Hsin Lin, Shuang Jia, and M.~Zahid
  Hasan.
\newblock Discovery of a weyl fermion semimetal and topological fermi arcs.
\newblock {\em Science}, 349(6248):613--617, 2015.

\bibitem{Xue1501092}
Su-Yang Xu, Ilya Belopolski, Daniel~S. Sanchez, Chenglong Zhang, Guoqing Chang,
  Cheng Guo, Guang Bian, Zhujun Yuan, Hong Lu, Tay-Rong Chang, Pavel~P.
  Shibayev, Mykhailo~L. Prokopovych, Nasser Alidoust, Hao Zheng, Chi-Cheng Lee,
  Shin-Ming Huang, Raman Sankar, Fangcheng Chou, Chuang-Han Hsu, Horng-Tay
  Jeng, Arun Bansil, Titus Neupert, Vladimir~N. Strocov, Hsin Lin, Shuang Jia,
  and M.~Zahid Hasan.
\newblock Experimental discovery of a topological weyl semimetal state in tap.
\newblock {\em Science Advances}, 1(10), 2015.

\bibitem{Liu2014}
Z.~K. Liu, J.~Jiang, B.~Zhou, Z.~J. Wang, Y.~Zhang, H.~M. Weng, D.~Prabhakaran,
  S.-K. Mo, H.~Peng, P.~Dudin, T.~Kim, M.~Hoesch, Z.~Fang, X.~Dai, Z.~X. Shen,
  D.~L. Feng, Z.~Hussain, and Y.~L. Chen.
\newblock A stable three-dimensional topological dirac semimetal cd3as2.
\newblock {\em Nat Mater}, 13(7):677--681, Jul 2014.
\newblock Letter.

\bibitem{RevModPhys.84.1067}
Valeri~N. Kotov, Bruno Uchoa, Vitor~M. Pereira, F.~Guinea, and A.~H.
  Castro~Neto.
\newblock Electron-electron interactions in graphene: Current status and
  perspectives.
\newblock {\em Rev. Mod. Phys.}, 84:1067--1125, Jul 2012.

\bibitem{Khveshchenko:2001zz}
D.~V. Khveshchenko.
\newblock {Ghost Excitonic Insulator Transition in Layered Graphite}.
\newblock {\em Phys. Rev. Lett.}, 87:246802, 2001.

\bibitem{Gamayun:2009em}
O.~V. Gamayun, E.~V. Gorbar, and V.~P. Gusynin.
\newblock {Gap generation and semimetal-insulator phase transition in
  graphene}.
\newblock {\em Phys. Rev.}, B81:075429, 2010.

\bibitem{Sabio:2010yf}
J.~Sabio, F.~Sols, and F.~Guinea.
\newblock {Variational approach to the excitonic phase transition in graphene}.
\newblock {\em Phys. Rev.}, B82:121413, 2010.

\bibitem{Montvay:1994cy}
I.~Montvay and G.~Munster.
\newblock {\em {Quantum fields on a lattice}}.
\newblock Cambridge University Press, 1997.

\bibitem{Drut:2008rg}
Joaquin~E. Drut and Timo~A. Lahde.
\newblock {Is graphene in vacuum an insulator?}
\newblock {\em Phys. Rev. Lett.}, 102:026802, 2009.

\bibitem{Hands:2008id}
Simon Hands and Costas Strouthos.
\newblock {Quantum Critical Behaviour in a Graphene-like Model}.
\newblock {\em Phys. Rev.}, B78:165423, 2008.

\bibitem{Armour:2009vj}
W.~Armour, Simon Hands, and Costas Strouthos.
\newblock {Monte Carlo Simulation of the Semimetal-Insulator Phase Transition
  in Monolayer Graphene}.
\newblock {\em Phys. Rev.}, B81:125105, 2010.

\bibitem{Drut:2009aj}
Joaquin~E. Drut and Timo~A. Lahde.
\newblock {Lattice field theory simulations of graphene}.
\newblock {\em Phys. Rev.}, B79:165425, 2009.

\bibitem{Ulybyshev:2013swa}
M.~V. Ulybyshev, P.~V. Buividovich, M.~I. Katsnelson, and M.~I. Polikarpov.
\newblock {Monte-Carlo study of the semimetal-insulator phase transition in
  monolayer graphene with realistic inter-electron interaction potential}.
\newblock {\em Phys. Rev. Lett.}, 111:056801, 2013.

\bibitem{Boyda:2016emg}
D.~L. Boyda, V.~V. Braguta, M.~I. Katsnelson, and M.~V. Ulybyshev.
\newblock {Many-body effects on graphene conductivity: Quantum Monte Carlo
  calculations}.
\newblock 2016.

\bibitem{Yamamoto:2016rfr}
Arata Yamamoto.
\newblock {Berry phase in lattice QCD}.
\newblock {\em Phys. Rev. Lett.}, 117(5):052001, 2016.

\bibitem{Yamamoto:2016zpx}
Arata Yamamoto and Taro Kimura.
\newblock {Quantum Monte Carlo simulation of topological phase transitions}.
\newblock {\em Phys. Rev.}, B94(24):245112, 2016.

\bibitem{Braguta:2016vhm}
V.~V. Braguta, M.~I. Katsnelson, A.~{\relax Yu}. Kotov, and A.~A. Nikolaev.
\newblock {Monte-Carlo study of Dirac semimetals phase diagram}.
\newblock 2016.

\bibitem{Scherer:2002tk}
Stefan Scherer.
\newblock {Introduction to chiral perturbation theory}.
\newblock {\em Adv. Nucl. Phys.}, 27:277, 2003.

\bibitem{Pich:1998xt}
Antonio Pich.
\newblock {Effective field theory: Course}.
\newblock In {\em {Probing the standard model of particle interactions.
  Proceedings, Summer School in Theoretical Physics, NATO Advanced Study
  Institute, 68th session, Les Houches, France, July 28-September 5, 1997. Pt.
  1, 2}}, pages 949--1049, 1998.

\bibitem{Bijnens:1995ww}
Johan Bijnens.
\newblock {Chiral Lagrangians and Nambu-Jona-Lasinio - like models}.
\newblock {\em Phys. Rept.}, 265:369--446, 1996.

\bibitem{Klevansky:1992qe}
S.~P. Klevansky.
\newblock {The Nambu-Jona-Lasinio model of quantum chromodynamics}.
\newblock {\em Rev. Mod. Phys.}, 64:649--708, 1992.

\bibitem{Susskind:1976jm}
Leonard Susskind.
\newblock {Lattice Fermions}.
\newblock {\em Phys. Rev.}, D16:3031--3039, 1977.

\bibitem{Braguta:2013rna}
V.~V. Braguta, S.~N. Valgushev, A.~A. Nikolaev, M.~I. Polikarpov, and M.~V.
  Ulybyshev.
\newblock {Interaction of static charges in graphene within Monte-Carlo
  simulation}.
\newblock {\em Phys. Rev.}, B89(19):195401, 2014.

\bibitem{Gockeler:1996me}
M.~Gockeler, R.~Horsley, V.~Linke, Paul E.~L. Rakow, G.~Schierholz, and
  H.~Stuben.
\newblock {Seeking the equation of state of noncompact lattice QED}.
\newblock {\em Nucl. Phys.}, B487:313--341, 1997.

\bibitem{Sekine:2014yna}
Akihiko Sekine and Kentaro Nomura.
\newblock {Stability of Multinode Dirac Semimetals against Strong Long-Range
  Correlations}.
\newblock {\em Phys. Rev.}, B90(7):075137, 2014.

\bibitem{Araki:2015php}
Yasufumi Araki.
\newblock {Lattice gauge theory treatment of strongly correlated Dirac
  semimetals}.
\newblock {\em PoS}, LATTICE2015:046, 2016.

\bibitem{Gonzalez:2015iba}
J.~Gonzalez.
\newblock {Phase diagram of the quantum electrodynamics of two-dimensional and
  three-dimensional Dirac semimetals}.
\newblock {\em Phys. Rev.}, B92(12):125115, 2015.

\bibitem{Gonzalez:2015tsa}
J.~Gonzalez.
\newblock {Strong-coupling phases of 3D Dirac and Weyl semimetals. A
  renormalization group approach}.
\newblock {\em JHEP}, 10:190, 2015.

\bibitem{2017arXiv170203551S}
A.~{Sharma}, V.~N. {Kotov}, and A.~H. {Castro Neto}.
\newblock {Excitonic Mass Gap in Uniaxially Strained Graphene}.
\newblock {\em ArXiv e-prints}, February 2017.

\end{thebibliography}

\end{document}